\documentstyle[psfig]{l-aa}

\def\kms{km~s$^{-1}$}
\def\vt{$v_{\rm t}$}
\def\teff{$T_{\rm eff}$}
\def\C12C13{$^{12}$C/$^{13}$C}
\begin{document}
\input{psfig.tex}

\thesaurus{
(08.01.1; 08.01.03; 08.16.3; 10.07.3)} 
\title{ Abundances for Globular Cluster Giants: }

\subtitle{ I. Homogeneous Metallicities for 24 Clusters }

\author{ E. Carretta\inst{1,3}, R.\ts G. Gratton\inst{2} }

\offprints{ E. Carretta. Table 9 also available in electronic form at the 
CDS via anonymous ftp 130.79.128.5 }

\institute{\inst{1} Osservatorio Astronomico di Bologna, Via Zamboni 33,
               I-40126 Bologna, ITALY\\
           \inst{2} Osservatorio Astronomico di Padova, Vicolo 
                    dell'Osservatorio 5, I-35122 Padova, ITALY\\
           \inst{3} Dipartimento d'Astronomia, Universit\`a di Padova,
              Vicolo dell'Osservatorio 5, I-35122 Padova, ITALY}

\date{Received 28 February 1996/Accepted 21 May 1996}

\maketitle

\begin{abstract}

We have obtained high resolution, high signal-to-noise ratio CCD echelle 
spectra of 10 bright red giants in 3 globular clusters (47 Tuc, NGC 6752 and 
NGC 6397) roughly spanning the whole range of metallicities of the galactic
globular cluster system. The analysis of this newly acquired material reveals 
no significant evidence of star-to-star variation of the [Fe/H] ratio in these
three clusters.
Moreover, a large set of high quality literature data (equivalent widths 
from high
dispersion CCD spectra) was re-analyzed in an homogeneous and
self-consistent way to integrate our observations and derive new metal
abundances for more than 160 bright red giants in 24 globular clusters ($i.e.$
about 16$\%$ of the known population of galactic globulars).
This set was then used to define a new metallicity scale for globular clusters
which is the result of high quality, direct spectroscopic data, of new and
updated model atmospheres from the grid of Kurucz (1992), and of a careful
fine abundance analysis; this last, in turn, is based on a common set of both
atomic and atmospheric parameters for all the stars examined.
Given the very high degree of internal homogeneity, our new scale 
supersedes the offsets and discrepancies existing in previous attempts to 
obtain a metallicity scale.
The internal uncertainty in [Fe/H] is very small: 0.06~dex (24
clusters) on average, and can be interpreted also as the mean precision of the cluster
ranking.
Compared to our system, metallicities on the widely used Zinn and West's scale
are about 0.10~dex higher for [Fe/H]$>-1$, 0.23~dex lower for $-1<$[Fe/H]$<-1.9$
and 0.11~dex too high for [Fe/H]$<-1.9$. The non-linearity of the Zinn and West's
scale is significant even at 3$\sigma$ level. A quadratic transformation is
given to correct older values to the new scale in the range of our calibrating
clusters ($-2.24 \le$[Fe/H]$_{ZW} \le -0.51$).
A minor disagreement is found at low metallicities between the metallicity
scale based on field and cluster RR Lyrae variables ($via$ a new calibration of
the $\Delta$S index) and our new cluster metallicities. It could be
tentatively ascribed to non-linearity in the [Fe/H]$-\Delta$S relationship.
The impact of new metallicities on major astrophysical problems is exemplified
through a simple exercise on the Oosterhoff effect in the classical pair M 3
and M 15.

\end{abstract}

%

\section{Introduction}

It is widely recognized that globular clusters (GCs) are cornerstones for the
solution of a large variety of problems concerning the formation and evolution
of galaxies. They are among the oldest objects formed in the Galaxy and, as a
consequence, can be used as tracers of the early chemical evolutionary phases
of the galactic environment. In principle, as a first guess, one can consider
the total metal abundance (traditionally indicated by the ratio 
[Fe/H]\footnote{We adopt the usual spectroscopic notation, $i.e.$ 
[X]= log(X)$_{\rm star} -$ log(X)$_\odot$ for any abundance quantity X, and 
log $\epsilon$(X) = log (N$_{\rm X}$/N$_{\rm H}$) + 12.0 for absolute number
density abundances.}) of a
globular cluster as representative of the original composition of the gas from
which it formed, and the variations in [Fe/H] among clusters as a fossil
recording of the chemical enrichment history occurred in the Galaxy. It is
nowadays well known, however, that this simple view is complicated by other
phenomena: the chemical composition of the atmospheres of stars can be altered
by processes occurring during their evolution (see Smith 1987, and Kraft 1994
for reviews). Even if the actual mechanisms are not completely understood yet,
it seems that surface abundances of only the lightest element (C, N, O, Na, and
Al) are affected, while Fe abundances are unchanged. In a following paper we
will discuss in detail this aspect. 

In the present paper we explore the possibility of building a new metallicity
scale for galactic globular clusters using only [Fe/H] values obtained from
fine analysis of high dispersion spectra. Although globular cluster distances
restrict applicability of this technique to the brightest giants, it provides
direct accurate and quantitative determination of the actual metallicity of
stellar atmospheres. 

It is rather surprising that despite the advent of CCD detectors and of
sophisticated and efficient spectrographs, the increasing number of measures of
cluster metallicities has been (and is!) almost totally ignored in a variety of
astrophysical problems involving this parameter. The most widely used
metallicity scale for globular clusters is in fact the one 
obtained by Zinn \& West
(1984; hereinafter ZW) and Zinn (1985) from a calibration of integrated
parameters of globular clusters. The main advantage of using integrated
parameters is that they can be easily measured even for the most distant
objects in the Galaxy: homogeneous results can then be obtained for almost all
known galactic clusters. However, integrated parameters are not directly
related to metal abundances, and their use as abundance indices requires an
accurate calibration in terms of the actual content of [Fe/H]. Reflecting
uncertainties present at that epoch in abundances from high dispersion spectra,
ZW attributed very low weight to direct abundance determinations for globular
cluster stars when they constructed their metallicity scale for globular
clusters.

Since ZW work, a number of accurate high-resolution spectroscopic
determinations of metal abundances for stars in globular clusters appeared in
the literature. However, only the work of Gratton and coworkers (Gratton et al
1986, Gratton 1987, Gratton and Ortolani 1989: G86, G87, G89 respectively;
collectively G8689) was aimed at a systematic determination of abundances for
a large number of clusters (spectra for giants in 17 clusters were actually
analyzed). However, since completion of the 
Gratton and coworkers survey, there has
been significant progresses both in the analysis techniques and on observing
facilities. In fact the new 
Kurucz (1992, hereinafter K92) model atmospheres allow an
homogeneous comparison between solar and stellar abundances, a major drawback
of former analysis of abundances for globular cluster stars (see $e.g.$,
Leep et al. 1987). Furthermore, improvements in high resolution
spectrographs and detectors allow better spectra to be obtained for a larger
number of stars: a major contribution has been done by observations with the
Hamilton spectrograph at Lick by Kraft, Sneden and coworkers (Sneden et al.
1991; Kraft et al. 1992; Sneden et al. 1992;  Kraft et al. 1993; Sneden et al.
1994; Kraft et al. 1995; hereinafter, SKPL1, SKPL2, SKPL3, SKPL4, SKPL5, SKPL6 
and SKPL on the whole, for
brevity). Regretfully, a homogeneous abundance scale based on high-resolution
spectroscopic data does not exist yet, mainly due to inconsistencies in the
model atmospheres and in the atomic parameters adopted in the various
investigations. 
Moreover, the need for such an improved scale is continuously growing, due to
the high degree of accuracy required by a variety of problems, one for all: the 
long-debated calibration of the absolute magnitude of the horizontal branch in
terms of the metal abundance. Minor variations in the adopted metallicities 
could result, ultimately, in non-trivial changes in globular 
clusters ages.

Our goal is to exploit recent observational and theoretical progresses to
construct a new, reliable metallicity scale for globular clusters, completely
based on high-quality data for red giant stars in 24 clusters. To this
purpose, we obtained new data for a few stars in three southern clusters using
the Long Camera mounted on the ESO CASPEC spectrograph, and reanalyzed published
equivalent widths ($EW$s), mainly from the Gratton and coworkers and Lick 
surveys, integrated by other sources of similar quality for clusters not 
included in those studies.
These data were analyzed in a totally self-consistent way, allowing a modern
calibration of the abundance indices considered by ZW. 

In Section 2 we present the new data, and in Section 3 the data 
adopted from literature. In
Sections 4 and 5 we discuss respectively the atmospheric and 
atomic parameters required for
the abundance analysis. Our results will be exposed in Section 6,
together with a discussion of the error sources. After a comparison with
previous works (Section 7), we present our conclusions in Section 8. 

\begin{table}
\begin{center}
\caption{ Observed spectral regions}
\begin{tabular}{lcl}
\hline
\hline
\\
$\lambda_c$~(\AA) & Spectral interv. & Indicators \\
\\ 
\hline
\\
4200 & 3800--4600 & CH (G band)       \\
5100 & 4700--5480 & C$_2$             \\
6300 & 5950--6700 & [O I] red doublet \\
8000 & 7750--8380 & CN red system     \\
\\
\hline
\end{tabular}
\label{t:regions}
\end{center}
\end{table}

\begin{table}
\begin{center}
\caption{ Program stars observed in 47 Tuc, NGC 6397 and NGC 6752}
\begin{tabular}{llccc}
\hline\hline
\\
Cluster  & Star  &$V$ & $B-V$ & $(V-K)_0$ \\
	 &       &    &       &           \\
\hline
\\
47~TUC   &  5422   &  12.47 & 1.40 & 3.23 \\
	 &  5529   &  11.87 & 1.59 & 3.86 \\
	 &  8406   &  12.37 & 1.43 & 3.33 \\
NGC~6397 &  C25    &  12.22 & 0.96 & 2.29 \\
	 &  C211   &  10.16 & 1.46 & 3.13 \\
	 &  C428   &  11.50 & 1.05 & 2.47 \\
NGC~6752 &  A31    &  10.80 & 1.60 & 3.69 \\
	 &  A45    &  11.57 & 1.23 & 3.00 \\
	 &  A61    &  11.71 & 1.13 & 2.91 \\
	 &  C9     &  12.37 & 1.03 & 2.62 \\
\\
\hline
\end{tabular}
\\
References - 47 Tuc: Frogel et al. (1981); NGC 6397 and NGC 6752: Frogel et al.
(1983). 
\label{t:program}
\end{center}
\end{table}

\section{Observational data}

We selected a sample of 10 red giant branch (RGB) stars in 3 GCs,
representative of the typical range of metal abundance of these objects, $i.e.$
47 Tuc (high metallicity, [Fe/H]$\simeq -0.8$~dex, 3 stars), NGC 6752
(intermediate metallicity, [Fe/H]$\simeq -1.5$, 4 stars) and NGC 6397 (low
metallicity, [Fe/H]$\simeq -2.0$, 3 stars). 

Due to obvious flux limitations, we restricted our observations to the
brightest globular cluster stars to obtain observational material good enough
($i.e.$ adequately high resolution and S/N ratios) for fine abundance analysis
with reasonable exposure times. All observed stars were 
brighter than V=12.5; the faintest giant observed, star 5422 in 47 Tuc, has M$_{\rm
V}^0 =-0.96$ (adopting a true distance modulus of 13.31 and A$_V$ = 0.12 from
Djorgovski 1993). Moreover, stars were selected to have infrared photometry
available, in particular in the K band, for an accurate determination of the
effective temperature \teff. 

The observational material was acquired in two runs: October 1990 and June
1991. In both runs, echelle spectra of the program stars were obtained with the
CASPEC spectrograph (in the Long Camera configuration) at the 3.6m ESO
telescope at La Silla, using a 31.6 lines/mm echelle grating. The slit
width was adjusted in order to give resolution $R\sim 30,000$. 

We tried to obtain a spectral coverage as large as possible, to observe
spectral features of different atomic and molecular species and to compare
abundances of the same element as derived from different indicators. In
Table~\ref{t:regions} we list the main features (abundance indicators)
of the spectral regions
observed. Table~\ref{t:program} lists the literature
photometric measurements of the program stars, and Table 3 the observed 
intervals for each star.

Exposure times ranged from 15 to 70 minutes to reduce cosmic ray contamination;
we usually tried to obtain more than one spectrum for each object to eliminate
the spurious events. A quartz lamp (for flat fielding) and a Thorium-Argon lamp
(for wavelength calibration) were acquired after each program star 
exposure, with
the telescope at the same position of the program star exposures. Besides,
fast-rotating early-type stars were observed each night to remove
telluric lines (see below). Bias frames have been taken at the beginning of
each night to account for readout noise. 

\begin{table}
\begin{center}
\caption{ Spectral intervals observed for the program stars}
\begin{tabular}{llcrr}
\hline\hline
\\
Cluster & Star & $\lambda_C$ &$t_{\rm exp}$&$S/N$\\
	&      &    (\AA)    & (min.)  &     \\
\\
\hline
\\
47 Tuc   & 5422 & 5100 & 240 & 145 \\
	 &      & 6300 & 100 & 111 \\
	 &      & 8000 & ~59 & ~45 \\
	 & 5529 & 4200 & ~70 & ~60 \\
	 &      & 5100 & ~70 & 100 \\
	 &      & 6300 & ~35 & ~56 \\
	 &      & 8000 & 120 & 100 \\
	 & 8406 & 5100 & 240 & 120 \\
	 &      & 6300 & ~90 & 107 \\
NGC 6397 & C25  & 6300 & ~80 & ~88 \\
	 & C211 & 4200 & ~40 & ~50 \\
	 &      & 6300 & ~15 & 117 \\
	 & C428 & 6300 & ~41 & ~91 \\
NGC 6752 & A31  & 6300 & ~30 & 130 \\
	 & A45  & 6300 & ~45 & 105 \\
	 & A61  & 6300 & ~50 & 101 \\
	 & C9   & 6300 & ~90 & 104 \\
\\
\hline
\end{tabular}
\label{t:coadded}
\end{center}
\end{table}

\subsection{Data analysis and equivalent widths}

The first steps in CCD reduction (bias subtraction, echelle order
identification, scattered light subtraction, order extraction and wavelength
calibration) were performed using standard packages implemented in
IRAF\footnote{IRAF is distributed by the National Optical Astronomy
Observatories, which is operated by the Association of Universities for
Research in Astronomy, Inc., under cooperative agreement with the National
Science Foundation.} environment. Off-order scattered light was eliminated 
through
bi-dimensional fitting along the dispersion and in the orthogonal
direction. The spectra were then wavelength calibrated using a dispersion
solution in two dimensions, derived from the Th-Ar lamps taken after each
spectrum, and one-dimensional spectra were extracted using an {\it optimal
extraction} algorithm implemented in the package.

The next steps of the analysis were then performed using the ISA package 
(Gratton 1988), purposely developed to deal with high-resolution spectra.

The blaze function was taken into account by dividing the spectra by dome flat
fields; the continuum was then traced on each individual spectrum for every object.
Whenever we had multiple exposures of the same star, spurious events and spikes
due to cosmic rays were eliminated comparing different spectra. 
We then used the spectra of featureless, rapidly rotating, early-type stars for
accurate removal of the telluric O$_2$ features, affecting in particular the
6,300~\AA\ region. We first identified atmospheric features in the solar 
spectrum then we measured each line in the spectra of early-type stars
acquired at different airmasses $z$. From these measurements we derived a mean
relationship between the $EW$s and $z$. For each star a synthetic spectrum of
the atmospheric lines was then computed and convolved with the instrumental
profile; O$_2$ features were finally cancelled out by dividing the spectrum
of each program star by the appropriate synthetic spectrum. This procedure 
allows
to correct each star for the appropriate airmass; moreover it does not 
introduce additional noise in the object spectra.

\begin{figure}[htbp]
\psfig{figure=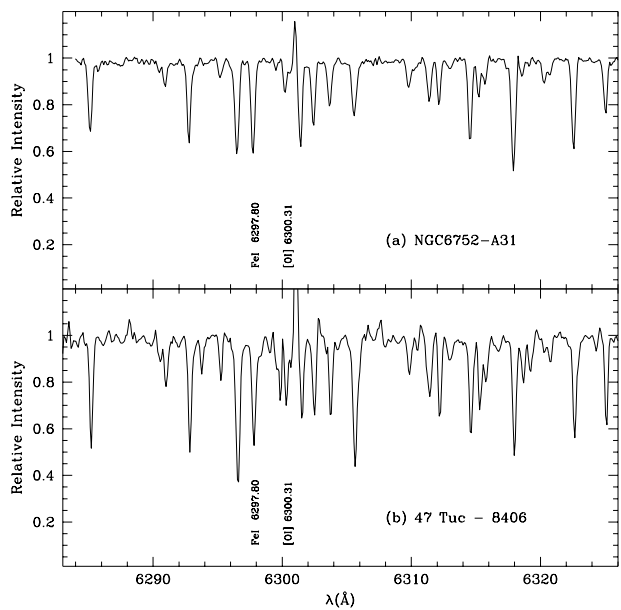,width=8.8cm,clip=}
\caption{ Tracing of a portion of the co-added, normalized spectra of 2 
stars in the 6,300~\AA\ region.}
\label{f:fig1}
\end{figure}

The resulting
cleaned spectra were then added to improve the $S/N$, after correcting for the
change in the (geocentric) radial velocity of the star. Finally, a new
continuum was traced. Figure~\ref{f:fig1} shows the tracing of a portion of 
our co-added and normalized CASPEC spectra
for two program stars.

Final available spectra are listed in Table~\ref{t:coadded}. We note that the
following abundance analysis uses only $EW$s measured in the spectral
regions centered at 5,100 and 6,300~\AA. The main reason is that they are less
affected by telluric bands and richer of stellar features (as compared to the
8,000~\AA\ region), and there is less concern related to line crowding and
continuum level identification (as compared to the 4,200~\AA\ region); this
last feature is of particular importance for the the spectra of stars in the
metal-rich cluster 47~Tuc.

\begin{table*}
{\tiny
\caption{Equivalent widths for giants in 47 Tuc, NGC 6397 and NGC 6752}
\begin{tabular}{lrrrrrrrrrrrrrr}
\hline\hline
\\
Element&$\lambda$(\AA)&E.P.&log $gf$&47Tuc&47Tuc&47Tuc& NGC6397 & NGC6397&NGC6397& NGC6752& NGC6752& NGC6752& NGC6752\\
       &         &      &       &  5422 &  5529 &  8406 &    25 &  C211 &  C428 &   A31 &   A45 &   A61 &    C9 \\
       &         &      &       &    EW &    EW &    EW &    EW &    EW &    EW &    EW &    EW &    EW &    EW \\
\\
\hline
\\
 Fe~I  & 5217.30 & 3.21 & -1.07 & 138.6 &   0.0 &   0.0 &   0.0 &   0.0 &   0.0 &   0.0 &   0.0 &   0.0 &   0.0 \\
 Fe~I  & 5250.21 & 0.12 & -4.90 &   0.0 &   0.0 & 197.0 &   0.0 &   0.0 &   0.0 &   0.0 &   0.0 &   0.0 &   0.0 \\
 Fe~I  & 5253.02 & 2.28 & -3.81 &   0.0 &   0.0 &  85.9 &   0.0 &   0.0 &   0.0 &   0.0 &   0.0 &   0.0 &   0.0 \\
 Fe~I  & 5956.70 & 0.86 & -4.57 & 166.4 &   0.0 & 177.9 &  31.3 & 101.3 &  41.7 &   0.0 & 116.8 & 103.0 &   0.0 \\
 Fe~I  & 5976.79 & 3.94 & -1.29 &  92.7 &   0.0 & 101.0 &   0.0 &  40.9 &  24.9 &  70.1 &  55.0 &  38.7 &   0.0 \\
 Fe~I  & 5984.82 & 4.73 & -0.28 &   0.0 &   0.0 &   0.0 &   0.0 &  35.0 &  17.7 &   0.0 &   0.0 &  42.3 &   0.0 \\
 Fe~I  & 6015.24 & 2.22 & -4.66 &  33.5 &   0.0 &  49.9 &   0.0 &   0.0 &   0.0 &  21.0 &   0.0 &   0.0 &   0.0 \\
 Fe~I  & 6019.37 & 3.57 & -3.21 &  15.4 &  24.9 &  15.8 &   0.0 &   0.0 &   0.0 &   0.0 &   0.0 &   0.0 &   0.0 \\
 Fe~I  & 6027.06 & 4.07 & -1.18 &  92.6 &   0.0 &  94.7 &   0.0 &  29.7 &   0.0 &  69.0 &   0.0 &   0.0 &   0.0 \\
 Fe~I  & 6034.04 & 4.31 & -2.35 &  14.0 &   0.0 &  27.6 &   0.0 &   0.0 &   0.0 &   0.0 &   0.0 &   0.0 &   0.0 \\
 Fe~I  & 6054.08 & 4.37 & -2.22 &  21.5 &   0.0 &   0.0 &   0.0 &   0.0 &   0.0 &   0.0 &   0.0 &   0.0 &   0.0 \\
 Fe~I  & 6056.01 & 4.73 & -0.45 &  64.6 &   0.0 &  81.1 &   0.0 &  18.4 &  11.3 &  52.8 &  43.1 &  37.9 &   0.0 \\
 Fe~I  & 6078.50 & 4.79 & -0.37 &  76.4 &   0.0 &  81.4 &   0.0 &   0.0 &   0.0 &   0.0 &   0.0 &   0.0 &  43.9 \\
 Fe~I  & 6079.01 & 4.65 & -0.95 &  54.9 &   0.0 &  56.5 &   0.0 &   0.0 &   0.0 &  22.7 &  32.6 &   0.0 &   0.0 \\
 Fe~I  & 6082.72 & 2.22 & -3.59 &   0.0 &   0.0 &   0.0 &   0.0 &  40.4 &   0.0 &   0.0 &  62.7 &   0.0 &   0.0 \\
 Fe~I  & 6093.65 & 4.61 & -1.29 &  33.3 &   0.0 &  41.2 &   0.0 &   0.0 &   0.0 &   0.0 &   0.0 &   0.0 &   0.0 \\
 Fe~I  & 6094.38 & 4.65 & -1.52 &   0.0 &   0.0 &  17.7 &   0.0 &   0.0 &   0.0 &   0.0 &   0.0 &   0.0 &   0.0 \\
 Fe~I  & 6096.67 & 3.98 & -1.73 &  53.5 &   0.0 &  82.3 &   0.0 &   0.0 &   0.0 &  41.7 &  29.8 &   0.0 &   0.0 \\
 Fe~I  & 6098.25 & 4.56 & -1.77 &   0.0 &   0.0 &  41.9 &   0.0 &   0.0 &   0.0 &   0.0 &   0.0 &   0.0 &   0.0 \\
 Fe~I  & 6120.26 & 0.91 & -5.94 &  76.0 &  95.4 &  78.1 &   0.0 &   0.0 &   0.0 &  64.8 &  32.5 &   0.0 &   0.0 \\
 Fe~I  & 6151.62 & 2.18 & -3.26 & 124.6 & 133.5 & 130.2 &   0.0 &  59.6 &  23.0 & 125.2 &   0.0 &  57.4 &  57.7 \\
 Fe~I  & 6157.73 & 4.07 & -1.25 &   0.0 &   0.0 & 107.6 &   0.0 &  34.5 &   0.0 &  78.8 &   0.0 &  48.5 &   0.0 \\
 Fe~I  & 6165.36 & 4.14 & -1.46 &  64.5 &  83.1 &  71.3 &   0.0 &   0.0 &   0.0 &  46.4 &  29.7 &   0.0 &  13.8 \\
 Fe~I  & 6173.34 & 2.22 & -2.85 & 145.9 &   0.0 & 147.1 &   0.0 &  82.6 &  46.6 & 157.7 &  94.2 &  91.5 &  87.2 \\
 Fe~I  & 6187.40 & 2.83 & -4.11 &  33.2 &   0.0 &   0.0 &   0.0 &   0.0 &   0.0 &   0.0 &   0.0 &   0.0 &   0.0 \\
 Fe~I  & 6187.99 & 3.94 & -1.58 &  69.6 &   0.0 &  83.0 &   0.0 &   0.0 &   0.0 &  47.7 &   0.0 &   0.0 &  23.5 \\
 Fe~I  & 6213.44 & 2.22 & -2.55 & 157.8 &   0.0 & 184.6 &  38.8 & 103.4 &  54.9 & 167.5 & 112.7 & 116.3 &  95.6 \\
 Fe~I  & 6219.29 & 2.20 & -2.42 & 170.2 & 184.3 & 180.0 &  62.8 & 115.8 &  66.5 & 187.9 & 134.2 & 131.1 & 116.3 \\
 Fe~I  & 6226.74 & 3.88 & -2.02 &   0.0 &   0.0 &  83.1 &   0.0 &   0.0 &   0.0 &  29.6 &   0.0 &   0.0 &   0.0 \\
 Fe~I  & 6240.65 & 2.22 & -3.23 &   0.0 &   0.0 & 137.0 &   0.0 &   0.0 &   0.0 &   0.0 &   0.0 &   0.0 &   0.0 \\
 Fe~I  & 6280.62 & 0.86 & -4.35 & 172.1 &   0.0 & 203.4 &   0.0 &   0.0 &  56.7 &   0.0 &   0.0 &   0.0 &   0.0 \\
 Fe~I  & 6297.80 & 2.22 & -2.73 & 170.9 &   0.0 & 180.5 &  38.6 &  99.1 &  50.6 & 152.6 & 113.2 & 112.8 &  90.6 \\
 Fe~I  & 6301.51 & 3.65 & -0.72 &   0.0 &   0.0 &   0.0 &  55.7 &   0.0 &   0.0 & 134.6 &   0.0 &   0.0 &   0.0 \\
 Fe~I  & 6311.51 & 2.83 & -3.14 &   0.0 &   0.0 &   0.0 &   0.0 &   0.0 &   0.0 &  72.9 &  30.9 &  21.7 &   0.0 \\
 Fe~I  & 6315.82 & 4.07 & -1.65 &  80.4 &   0.0 &  90.5 &   0.0 &   0.0 &   0.0 &   0.0 &  37.6 &   0.0 &   0.0 \\
 Fe~I  & 6322.69 & 2.59 & -2.39 & 135.3 &   0.0 & 149.2 &   0.0 &  79.3 &   0.0 & 141.8 & 104.8 &  93.7 &  72.6 \\
 Fe~I  & 6353.84 & 0.91 & -6.44 &  53.6 &   0.0 &   0.0 &   0.0 &   0.0 &   0.0 &   0.0 &   0.0 &   0.0 &   0.0 \\
 Fe~I  & 6380.75 & 4.19 & -1.32 &  78.7 &   0.0 &  90.3 &   0.0 &  16.5 &  15.2 &  52.8 &  40.0 &  22.6 &  34.0 \\
 Fe~I  & 6392.54 & 2.28 & -3.95 &  74.3 &  74.2 &  69.8 &   0.0 &   0.0 &   0.0 &   0.0 &  25.9 &  21.7 &   0.0 \\
 Fe~I  & 6393.61 & 2.43 & -1.43 & 216.1 &   0.0 &   0.0 &  87.2 &   0.0 &  99.6 & 222.5 &   0.0 &   0.0 &   0.0 \\
 Fe~I  & 6498.95 & 0.96 & -4.66 & 151.3 &   0.0 &   0.0 &   0.0 &  86.2 &   0.0 & 172.1 &  97.4 &  81.8 &  77.5 \\
 Fe~I  & 6518.37 & 2.83 & -2.46 & 113.3 &   0.0 & 120.6 &  17.3 &  47.4 &  22.6 & 119.7 &  72.6 &  59.4 &  46.3 \\
 Fe~I  & 6581.22 & 1.48 & -4.68 & 108.8 &   0.0 &   0.0 &   0.0 &  32.8 &   0.0 & 118.1 &  48.5 &  36.8 &   0.0 \\
 Fe~I  & 6593.88 & 2.43 & -2.39 & 159.6 &   0.0 & 158.0 &   0.0 & 106.0 &  49.7 &   0.0 & 124.4 & 118.6 & 104.9 \\
 Fe~I  & 6608.03 & 2.28 & -3.94 &  95.5 &   0.0 &  95.7 &   0.0 &   0.0 &   0.0 &  65.2 &  26.5 &  27.2 &  18.4 \\
 Fe~I  & 6609.12 & 2.56 & -2.66 & 136.8 &   0.0 & 144.0 &  16.5 &  63.4 &  25.3 & 134.7 &  89.1 &  74.7 &  55.0 \\
 Fe~I  & 6667.43 & 2.45 & -4.35 &   0.0 &   0.0 &   0.0 &   0.0 &   0.0 &   0.0 &  17.8 &   0.0 &   0.0 &   0.0 \\
 Fe~II & 5264.81 & 3.23 & -3.21 &  43.7 &   0.0 &   0.0 &   0.0 &   0.0 &   0.0 &   0.0 &   0.0 &   0.0 &   0.0 \\
 Fe~II & 5991.36 & 3.15 & -3.55 &   0.0 &   0.0 &   0.0 &  15.0 &  19.3 &   0.0 &  29.5 &  29.6 &  30.4 &   0.0 \\
 Fe~II & 6084.11 & 3.20 & -3.80 &   0.0 &   0.0 &   0.0 &   0.0 &  17.8 &   0.0 &  20.4 &   0.0 &   0.0 &   0.0 \\
 Fe~II & 6149.25 & 3.89 & -2.73 &   0.0 &   0.0 &   0.0 &   0.0 &  14.0 &   0.0 &   0.0 &   0.0 &  30.9 &   0.0 \\
 Fe~II & 6369.46 & 2.89 & -4.21 &   0.0 &   0.0 &   0.0 &   0.0 &  14.0 &  10.8 &   0.0 &  20.1 &  13.5 &   0.0 \\
 Fe~II & 6432.68 & 2.89 & -3.58 &   0.0 &  30.0 &  30.0 &  16.7 &  43.7 &  25.7 &   0.0 &  38.1 &   0.0 &  34.3 \\
 Fe~II & 6456.39 & 3.90 & -2.09 &  52.1 &  48.0 &  44.2 &   0.0 &  49.9 &  37.8 &  48.1 &  52.0 &  57.9 &  48.0 \\
 Fe~II & 6516.09 & 2.89 & -3.38 &   0.0 &   0.0 &  45.9 &   0.0 &   0.0 &  45.3 &  49.2 &  48.3 &  57.0 &  38.6 \\
\\
\hline
\end{tabular}
\label{t:ew}
}
\end{table*}

Equivalent widhts of various element were then measured in the two mentioned
regions: in the following 
we will refer only to iron lines. Gaussian profiles were fitted
to the observed profiles; when the number of clean lines was very low ($e.g.$, for
Fe~II lines), we first derived a relationship between $EW$\ and central depth
$r_{\rm C}$ from unblended lines and then we used it to add some new
EW's for measured $r_{\rm C}$'s. The number of measured lines depends on 
$S/N$\ and on the star
metallicity; generally, some 25-50 Fe lines were measured for NGC 6397 stars, 
some 50-70 for NGC 6752 stars and from 50 to 100 for 47 Tuc stars. 
In the following analysis, only lines with $EW>10$~m\AA\ were
used. Line parameters (see Section 5) and $EW$s are listed in 
Table~\ref{t:ew}, both for Fe~I and Fe~II lines.

\begin{figure}[htbp]
\psfig{figure=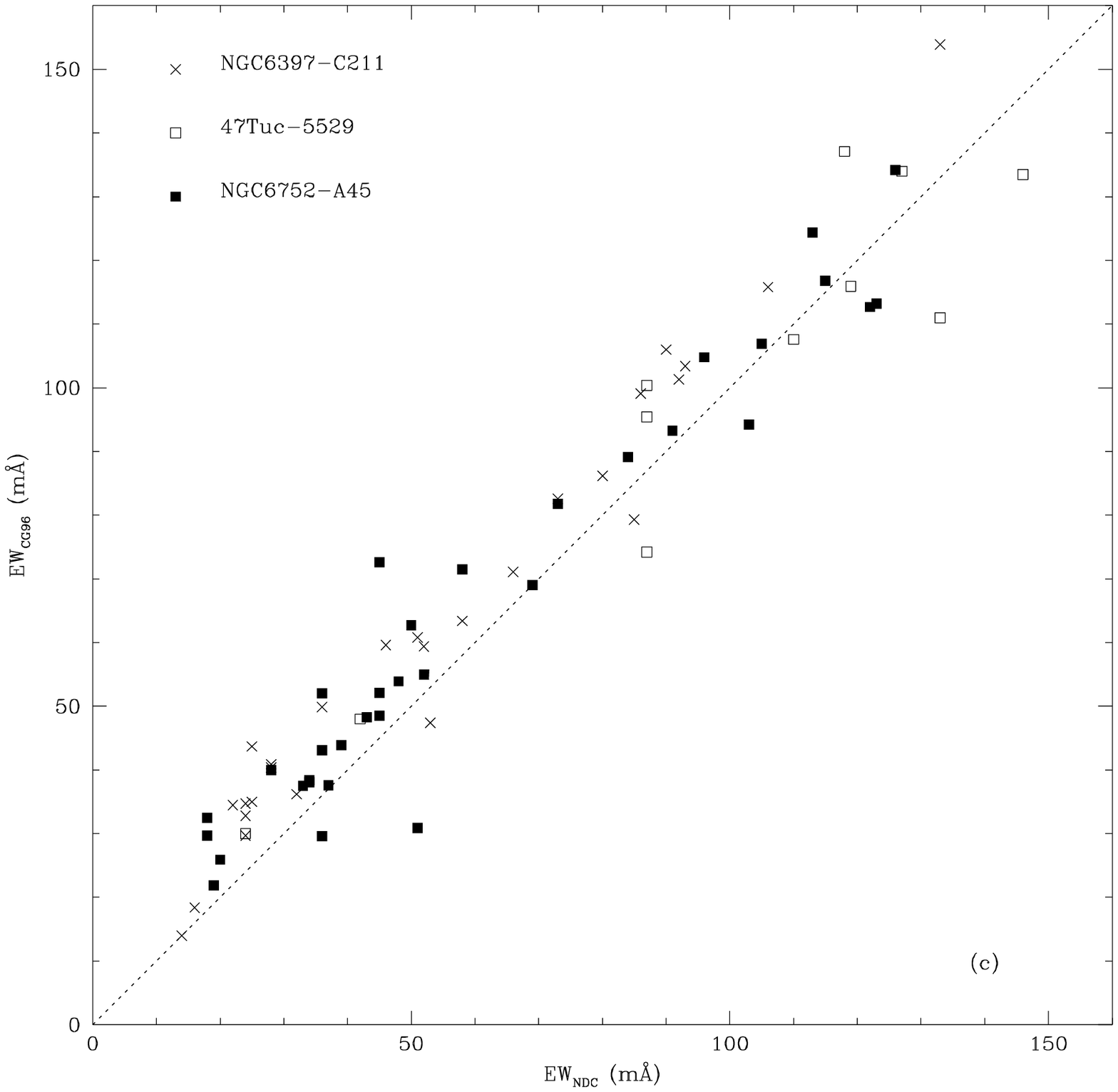,width=8.8cm,clip=}
\caption{ Comparison of our $EW$s with those of NDC for the 3 stars in common}
\label{f:fig2}
\end{figure}

We have 3 stars (namely, 47 Tuc-5529, NGC 6397-C211 and NGC 6752-A45) in common 
with another recent, high dispersion analysis by Norris and Da Costa (1995; 
hereinafter, NDC), at about the same resolution we used: this allows a
comparison between the two sets of $EW$s, which is shown in Figure~\ref{f:fig2}.
The average difference is\footnote{We will indicate with the subscript CG96 the
sample of our observed program stars and simply CG the results obtained using
the whole sample (observations plus literature data) analyzed in the present
paper}: $EW_{\rm CG96} - EW_{\rm NDC} = 6.1 \pm 1.1$~m\AA\ ($\sigma$=9.2~m\AA, 
76 lines\footnote{Throughout this paper, the symbol 
$\sigma$ will indicate the standard deviation of a single measurement, while 
the value after the
symbol $\pm$ will refer to the standard deviation of the mean}).
Note that here we regard this comparison as an external check of the accuracy
of our $EW$'s measurements, but since we used data from NDC to enlarge 
the sample of analyzed stars (see below), the above comparison has to be 
regarded also as a self-consistency test on our total set of $EW$s.
 
\section{Literature data}

To increase our sample and to obtain a more statistically significant basis for
our discussion, we also analyzed in a homogeneous way the globular cluster
giants studied in two recent surveys: G8689 (41 red giants in 17 globular
clusters) and SKPL (82 giants in 7 clusters). 
To these sets, we also added a full re-analysis of 3 giants in NGC 2298 
(from McWilliam et al. 1992; McW92), 18 stars studied by Minniti et al. 
(1993; M93) in 8 clusters, and 8 stars in 2 clusters from NDC.
Table~\ref{t:comp} lists the
whole sample of globular clusters red giants analyzed in the present paper,
with the source for each subset of data.

\begin{table}
\begin{center}
\caption{ Complete list of globular clusters analyzed}
\label{t:comp}
\begin{tabular}{lrr}
\hline\hline
\\
GC &Nr.Stars & Reference \\
\\
\hline
\\
NGC 104-47 Tuc &  3 & CG96  \\
               &  2 & G8689 \\
NGC 288        &  2 & G8689 \\
NGC 362        &  2 & G8689 \\
NGC 1904       &  2 & G8689 \\
NGC 2298       &  3 & McW92 \\
NGC 3201       &  3 & G8689 \\
NGC 4590-M 68  &  2 & G8689 \\
               &  2 & M93   \\
NGC 4833       &  2 & G8689 \\
               &  1 & M93   \\
NGC 5272-M 3   & 10 & SKPL  \\
NGC 5897       &  2 & G8689 \\
NGC 5904-M 5   & 13 & SKPL  \\
               &  3 & G8689 \\
NGC 6121-M 4   &  3 & G8689 \\
NGC 6144       &  1 & M93   \\
NGC 6254-M 10  &  2 & G8689 \\
               & 14 & SKPL  \\
NGC 6205-M 13  & 23 & SKPL  \\
NGC 6341-M 92  &  9 & SKPL  \\
NGC 6352       &  3 & G8689 \\
NGC 6362       &  2 & G8689 \\
NGC 6397       &  3 & CG96  \\
               &  3 & G8689 \\
               &  5 & M93   \\
               &  2 & NDC   \\
NGC 6656-M 22  &  3 & G8689 \\
NGC 6752       &  3 & G8689 \\
               &  4 & CG96  \\
               &  6 & NDC   \\
               &  3 & M93   \\
NGC 6838-M 71  & 10 & SKPL  \\
               &  3 & G8689 \\
NGC 7078-M 15  &  3 & SKPL  \\
               &  2 & M93   \\
NGC 7099-M 30  &  2 & M93   \\
\\
\hline
\end{tabular}
\end{center}
\end{table}

\begin{figure}[htbp]
\psfig{figure=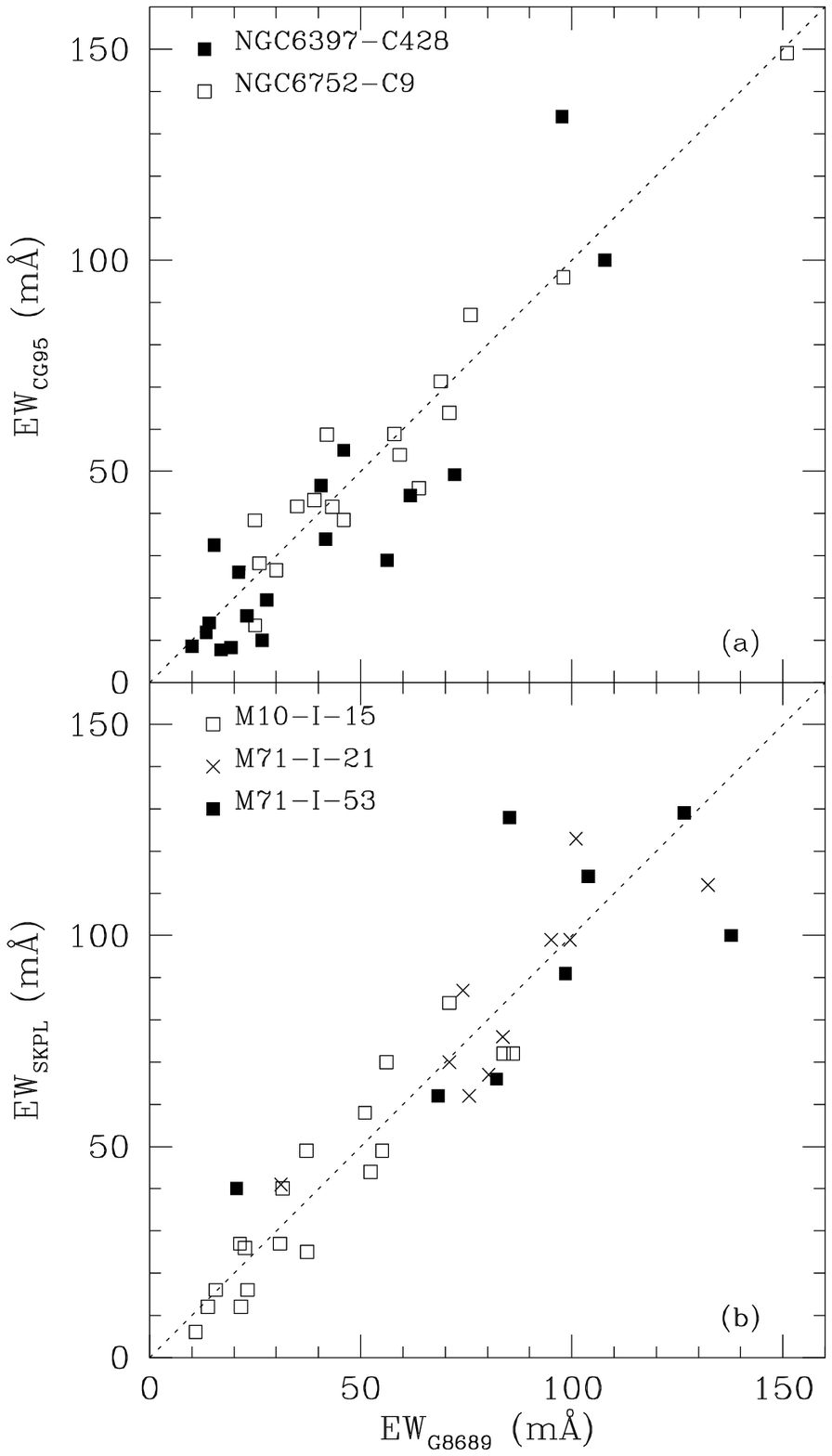,width=8.8cm,clip=}
\caption{(a) Comparison between $EW$s of G8689 sample after the correction and
the $EW$s of the present analysis (sample CG96) for NGC 6752-C9 and NGC
6397-C428; (b) another test of self-consistency of our $EW$s, done by comparing
the corrected $EW$s from G8689 sample with those from SKPL sample for 3 stars
in common between the two sets.} 
\label{f:fig3ab}
\end{figure}

We simply adopted published $EW$s for all samples, since the observational
material consists in high-resolution ($R\sim 30,000$), high $S/N$\ spectra,
fairly comparable to our own spectra. However, for the G8689 sample we had to
make some corrections to the published $EW$s, before using them for a new
analysis. In fact, in G87 the authors themselves noted that their $EW$s were
systematically higher than M\"ackle et al. (1975) $EW$s for $\alpha$ Boo
(Arcturus). This is due to the lower resolution (R $\sim 15,000$) used by
G8689: unnoticed lines blend with the measured
ones, resulting in an overall overestimate of the $EW$s. We derived the
following relation between G8689 $EW$s ($EW_{\rm G}$) and M\"ackle et al.
(1975) $EW$s ($EW_{\rm M}$) for $\alpha$~Boo : 
\begin{equation}
\label{eq:1}
EW_{\rm M} = 0.914\,(\pm 0.044) EW_{\rm G} - 14.3\,(\pm 17.6)~~{\rm m\AA}
\end{equation}
Strictly, this correction applies only to $EW$s measured on spectra having
metallicities similar to that of Arcturus ([Fe/H]$\sim -0.5$); we expect
smaller corrections for more metal-poor stars. To verify this point we
considered the stars in common between G8689 and the present data (NGC
6397-C428 and NGC 6752-C9); in both cases there are 17 lines in common between
the old and the new analyses. The mean difference between the $EW$s of our
new analysis $EW_{\rm new}$\ and those of G8689 for these two stars are: 
\begin{equation}
EW_{\rm new}-EW_{\rm G} = -3.0 \pm 1.8~~{\rm m\AA}
\end{equation}
($\sigma=10.5$~m\AA, 34 lines). Had we applied first eq.~(\ref{eq:1}) to
correct the G8689's EWs, this difference would have been: 
\begin{equation}
EW_{\rm new}-EW_{\rm G,corr} = 15.3 \pm 1.8~~{\rm m\AA}
\end{equation}
($\sigma=10.3$~m\AA, 34 lines), where $EW_{\rm G,corr}$\ are G8689 $EW$s after
transformation using eq.~(\ref{eq:1}). The overall correction is $\sim
18$~m\AA. We then estimate that while eq.~(\ref{eq:1}) gives the appropriate
corrections to G8689 $EW$s for metallicities similar to that of Arcturus
([Fe/H]$\sim -0.5$), these corrections should reduce to $\sim 1/6$\ of
this amount
for a metallicity in between those of NGC 6752 and NGC 6397 ([Fe/H]$\sim
-1.7$). We then included a metallicity-dependent correction, obtaining: 
\begin{equation}
\label{eq:4}
EW_{\rm C}=EW_{\rm G}-(0.7\,{\rm [Fe/H]}+1.35)[(1-a)\,EW_{\rm G}-b]
\end{equation}

Eq.~(\ref{eq:4}) was then applied to the old $EW_{\rm G}$s of G8689 to bring
them to the same system of our higher resolution spectra. 
Figure~\ref{f:fig3ab} (a)
compares $EW_{\rm C}$\ with $EW_{\rm new}$\ for the two stars in common between
the G8689 and the CG96 sample; the average difference is $EW_{\rm C}-EW_{\rm
new}=3\pm 2$~m\AA\ ($\sigma$ = 10~m\AA, 34 lines). The best fit line (with
constant term put to zero) to the data in Figure~\ref{f:fig3ab} (a) is: $EW_{\rm
C}=1.025\pm 0.035 EW_{\rm new}$, with a scatter of 11~m\AA. In the following
analisys, we used only lines with $EW_{\rm C}>$20~m\AA. 

Figure~\ref{f:fig3ab} (b) shows another internal comparison, using 
G8689 (corrected) and SKPL $EW$s for 3 stars: the agreement is very good, the
average difference being 
$EW_{\rm SKPL} - EW_{\rm G8689} = -0.5 \pm 2.4$~m\AA\ ($\sigma$=14~m\AA, 36 
lines). We estimated internal errors in $EW$s of about $\pm$9~m\AA\ in G8689
set and $\pm$5~m\AA\ in CG96, SKPL and the other sets analyzed.
We can then be confident that our results are based on an homogeneous and 
self-consistent dataset.
 
\section{Atmospheric parameters}

Model atmospheres appropriate for each star were extracted from the grid of
K92\footnote{ The model atmospheres considered in
this paper are those in Kurucz's CD-ROM 13. In these models, convection description
include an approximate consideration of possible overshooting in the stellar
atmospheres. This point is discussed at length in Castelli et al. (1996), who
conclude that at present is not possible to establish whether or not this treatment of
convection should be preferred to a more traditional approach without any
overshooting. However, this uncertainty is of only minor concern for the
metallicity scale established in this paper, insofar the approach used for the
Sun and the globular cluster giants is the same. }
Values of effective temperature \teff\ and surface gravity $\log g$\ for giants
observed in 47 Tuc, NGC 6752 and NGC 6397 (sample CG96) were taken from Frogel
et al. (1981, 1983). Frogel et al's \teff's are based on visual-near infrared
colours (mainly $V-K$), transformed into temperatures based on the Cohen et al.
(1978; hereinafter CFP) scale. $\log g$ values are determined from the position
of the stars in the colour-magnitude diagram (CMD). These values are considered
as accurate as $\pm$100~K in \teff\ and $\pm$0.3~dex in log $g$, including
possible uncertainties in the effective temperature scale, as well in the
adopted red giant mass, in the bolometric corrections and in the cluster
distance modulus. 

\teff\ and $\log g$\ values for stars in the SKPL sample are also generally based on
CFP's scale, except for giants in M 92 and M 15; in these two cases the authors
used \teff's from Carbon et al. (1982) and Trefzger et al. (1983).
The reader should be warned that final values for \teff\ and log $g$, in the
SKPL sample, are slightly different from the actual values on the
CFP's scale. This is because SKPL adjusted the adopted photometric
temperatures until a slope close to zero was achieved for the relationship
between Fe~I abundances from individual lines and excitation potential.
However, as discussed in their papers, the agreement of photometric and final
spectroscopic \teff\ was from the beginning 
well within the uncertainties of the temperature
scale itself, so that neither the abundance ratios for individual stars nor the
mean values for the clusters are significatively changed by adopting the
tabulated values of atmospheric parameters from SKPL.

In the G8689 sample, \teff\ and $\log g$ values were homogeneously adopted
from Frogel et al. (1979, 1981, 1983) and are also used in the present work.
For other data sets (McW92, NDC) we were able to use atmospheric parameters
from Frogel and coworkers. Only half of the 18 stars from M93 have \teff\
and $\log g$ values listed in Frogel et al.; for the missing stars, we used
their original values, derived spectroscopically from the dependence of 
\teff\
on excitation  potential and from the Fe ionization equilibrium. In fact, the
parameters adopted by M93 are with very good approximation on the CFP's scale: 
the mean for the stars in common are: 
T$_{\rm eff}$(Frogel)-T$_{\rm eff}$(M93)=$-29\pm11$~K
($\sigma$=30~K, 8 stars) and log $g$(Frogel)- log $g$(M93)=$-0.10\pm0.06$~dex
($\sigma$=0.17~dex, 7 stars). 
For further details, see the original papers of G8689 and SKPL. The
starting input values for [Fe/H] were those from the original analyses.

Metallicities were obtained varying the metal abundance [A/H] of the model
until it was equal to the derived [Fe/H] value. For the microturbulent 
velocity \vt, the input values were changed until no trend in Fe abundance 
with the $EW$\ of Fe~I lines was present \footnote{Expected line strengths
were used when determining the microturbulent velocity, following the prescription of 
Magain (1984)}. However, for giants in M 92 and M 15 only a few lines were 
available, so we took the values of \vt\ from the relationship: 
\begin{equation}
v_{\rm t} = -0.322 (\pm 0.048) \cdot \log g + 2.22 (\pm 0.31)
\end{equation}
derived elsewhere (Carretta, Gratton \& Sneden 1996; Gratton \& Carretta 1996)
for field stars. The adopted atmospheric parameters are listed in
Table 9 (also available in electronic form) for all the stars
studied in the present paper. Metallicities were
obtained varying the metal abundance [A/H] of the model until it was equal to
the derived [Fe/H] value. 

\section{Oscillator strengths, line selection and solar abundances}

Only lines clean from blends on very high resolution solar spectra were
considered in the analysis; for Fe, the line list was extracted from 
Rutten \& van der Zalm (1984), and Blackwell et al. (1980). The adopted values
for the oscillator strengths $gf$s were determined following the same precepts
of Clementini et al. (1995), who performed a high-dispersion study of the metal
abundances of field RR Lyrae variables. Briefly, whenever possible laboratory
$gf$s were considered: for Fe~I lines they were taken from papers of the Oxford
group (for references, see Simmons \& Blackwell, 1982) and Bard et al. (1991)
and Bard \& Koch
1994), $gf$s of the Oxford group being corrected upward to account for the
systematic difference with those of Bard et al. (0.03~dex; see Clementini et
al. 1995). $gf$s for Fe~II lines were taken from Heise \& Kock (1990),
Bi\'emont et al. (1991) and Hannaford et al. (1992). For lines lacking accurate
laboratory determinations, $gf$s were derived from an inverse solar analysis
using the Holweger and M\"uller (1974: HM) model atmosphere and the Fe
abundance derived from the other lines. The adopted $gf$\ values are also
listed in Table~\ref{t:ew}; they can be integrated for other lines using 
$gf$s in the larger line list of Clementini et al. 
(1995, tables 3(a),(b)).

In any differential analysis ($i.e.$ a comparative analysis in which the 
zero point of the [Fe/H] scale is set by the Sun) the assumed solar abundance
is of paramount importance. The solar Fe abundance obtained with the set of 
$gf$s described above and the K92 solar model
atmosphere ($\log \epsilon$(Fe)=7.52), taken as the reference value for the Sun
throughout the present work, agrees with the meteoritic value of Anders \&
Grevesse (1989: $\log \epsilon$(Fe)=7.51). It is also very close to the value
obtained using the HM model atmosphere ($\log \epsilon$(Fe)=7.56: Castelli
et al. 1996). A very
similar value is obtained using Fe~II lines when adopting the K92 model. The
trend with excitation potential is also very small in the Sun. 

We wish to stress here the importance of the consistency between these various
determinations of the solar Fe abundances: in fact, the whole scale of metal
abundance previously determined for globular clusters 
was uncertain, due to the rather large difference ($\sim 0.15$~dex) between the
solar Fe abundances obtained using the HM and the Bell et al. (1976: hereinafter
BEGN) model atmospheres generally adopted in the analysis of cool cluster
giants (see $e.g.$, Leep et al. 1987). In fact, it was not
clear what solar Fe abundance to use: either the one determined using a
model extracted from the same grid used for cluster giants, or that obtained
using the best solar model. Use of K92 atmospheres solves this problem, since we
may now use models for giants extracted from the same grid which gives a solar
Fe abundance in agreement with the best photospheric and meteoritic
determinations. 

Finally, we note that this choice of lines and of $gf$ values allows a direct
comparison with the metallicity scale for the low-dispersion index $\Delta$S\ in 
RR Lyrae stars derived by Clementini et al. (1995).
   
\begin{figure}[htb]
\psfig{figure=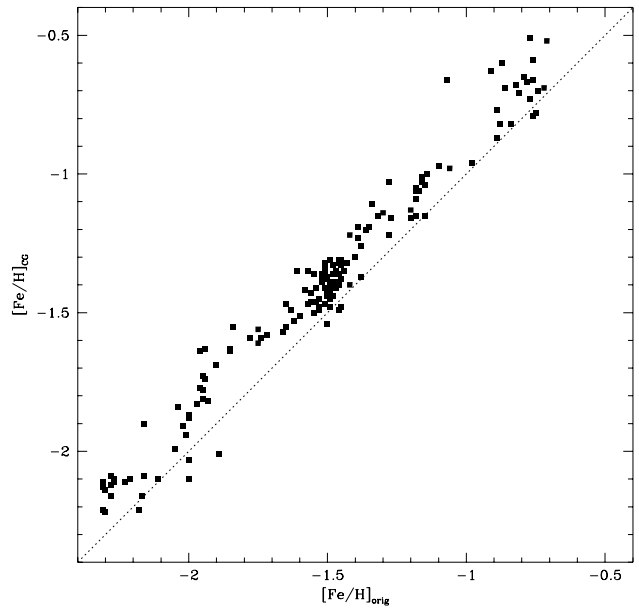,width=8.8cm,clip=}
\caption{ Comparison of new [Fe/H] values with those from the original
analyses. Different symbols represent stars of different samples studied.}
\label{f:fig4}
\end{figure}

\section{Results and error estimates}

Our Fe abundances for both the original program stars and for all reanalyzed
stars are listed in Table 9 (last 7 columns)\footnote{All stars 
that were in more than one data sets have been independently re-analyzed, and
then their [Fe/H] values averaged before computing the cluster's mean [Fe/H].}.
Final [Fe/H] values adopted (last column) are those derived from neutral
lines alone, since the number of Fe~II lines with accurate $EW$s was often too
small. When individual clusters are considered, our [Fe/H] values do not show 
any trend with \teff\ on the whole range 3800--4900~K (which approximatively
corresponds to a range of about 2.5 in M$_{\rm bol}$). 

Our [Fe/H] values are systematically higher than those of the original
analyses: the systematic difference is $0.12\pm 0.01$~dex ($\sigma$=0.08, 162
stars), as displayed also in Figure~\ref{f:fig4}. This difference is mainly due
to our use of K92 model atmospheres for both solar and stellar analysis.
In fact, in previous analyses ($e.g.$, both G8689 and SKPL) the solar Fe abundances
were obtained using the HM model atmospheres, which is $\sim 150$~K warmer than
the BEGN models in the line formation region. We notice here that relative
abundances ($i.e.$ abundances obtained using model atmospheres from the same grid
for both the Sun and the program stars) are almost insensitive to the grid
adopted (differences are $<0.03$~dex). In this respect, our analysis combines 
the advantages of both differential and absolute analyses, since our
abundances are referred to the Sun, and we used a solar model extracted from the
same grid of model atmospheres used for the program stars.

\begin{table}
\begin{center}
\caption{ Mean differences Fe~I - Fe~II in globular cluster giants }
\begin{tabular}{lrcc}
\hline\hline
\\
Sample & Nr. Stars & $<$Fe~I -- Fe~II$>$ &$\sigma$ \\
\\
\hline
\\
All  & 147 &  ~0.00 $\pm$ 0.01 & 0.12 \\
CG96 &  10 &$-$0.05 $\pm$ 0.04 & 0.11 \\
G8689&  29 &$-$0.10 $\pm$ 0.03 & 0.14 \\
SKPL &  82 &  +0.02 $\pm$ 0.01 & 0.08 \\
McW  &   3 &  +0.13 $\pm$ 0.04 & 0.07 \\
M93  &  15 &  +0.11 $\pm$ 0.04 & 0.15 \\
NDC  &   8 &$-$0.01 $\pm$ 0.02 & 0.06 \\ 
\\
\hline
\end{tabular}
\label{t:mean}
\end{center}
\end{table}

Data concerning the Fe ionization equilibrium are shown in Table~\ref{t:mean}
which lists the mean differences between abundances derived from neutral and
singly ionized lines of Fe. These values have been computed both for the
total sample and for the different sub-samples studied. From this Table we
conclude that there is an excellent agreement between abundances derived from
Fe~I and Fe~II, with no trend with \teff\ or [Fe/H]. The lack of any trend over
the whole range of temperature is very important, since in the past some
analysis ($e.g.$, Pilachowski et al. 1983) claimed that a discrepancy was present 
between these 2 iron abundances in stars cooler than 4,300~K. The implication
was that in the very upper red giant branch the usual Local Thermodynamic
Equilibrium (LTE) assumption had to be released or, at least, carefully
verified by statistical equilibrium computations. Our results, however, 
strongly confirms the recent study of Clementini et al. (1995; see also the 
footnote below) that pointed out that departures from LTE cannot 
significatively affect abundance analyses for stars cooler than RR Lyrae
variables.

For the SKPL sample it should be noted again that in their
original papers both photometric gravities and \teff\ values were
purposedly changed to obtain a match of the two [Fe/H] abundances within
0.05~dex. 

\begin{table*}
\begin{center}
\caption{ Dependence of the derived abundances on atmospheric parameters }
\begin{tabular}{lrrrrrrc}
\hline
\hline 
\\
 Element  &  $\Delta$\teff\  
 &$\Delta\log g$ &$\Delta{\rm [A/H]}$  &$\Delta$\vt 
 &$\Delta_{tot}$ &$\Delta_{tot}$ &$\Delta_{mod}$ \\
 &$+$100K  &$+$0.5 dex &$+$0.2 dex &$+$0.5 \kms &random  & syst.  & BEGN-K92\\
\\
\hline
\\
\multicolumn{8}{c}{Star 8406 in 47 Tuc (${\rm [A/H]}$ = -0.61)}
\\
${\rm [Fe/H]}$I &  0.015 & 0.118 & 0.044 & -0.232 & 0.100 & 0.080 & -0.124\\
${\rm [Fe/H]}$II& -0.185 & 0.318 & 0.091 & -0.094 & 0.136 & 0.183 & -0.165\\
\\
\hline
\\
\multicolumn{8}{c}{Star A61 in NGC 6752 (${\rm [A/H]}$ = $-$1.61)}
\\
${\rm [Fe/H]}$I & 0.149 & 0.017 & -0.015  & -0.094 & 0.082 & 0.079 & -0.039\\
${\rm [Fe/H]}$II&-0.077 & 0.235 &  0.058  & -0.053 & 0.081 & 0.117 & -0.017\\
\\
\hline
\end{tabular}
\label{t:error}
\end{center}
\end{table*}

Table~\ref{t:error} shows the dependance of the derived abundances from
uncertainties in the adopted atmospheric parameters; this is obtained by
re-iterating the analysis while varying each time only one of the parameters.
To show how these sensitivities change with overall metal abundance, we
repeated this exercise for both a metal-rich (star 8406 in 47~Tuc) and a
metal-poor giant (star A61 in NGC 6752). 

Entities of variations are quoted in Table~\ref{t:error}: these
values are larger than errors likely present in the adopted atmospheric
parameters. This will be shown in the following discussion, where we will try
to provide reasonable evaluations 
for the uncertainties in the adopted atmospheric
parameters. To this purpose, we compared expected scatters in Fe abundances
within individual clusters and differences between abundances provided
by neutral and singly ionized lines with observed values. Relevant data for 
this last parameter can be easily obtained from
Table~\ref{t:mean}\footnote{Following the non-LTE analysis of Clementini et
al. 1995, no significant departures from LTE are to be expected in RGB stars.
The differences found in Fe abundances from Fe~I and Fe~II lines are thus
likely to be interpreted as due to errors in the analysis and
in the atmospheric parameters.}. For the reasons above mentioned, we omit from
the following discussion the value from the SKPL sample and we concentrate
instead on the other mean differences, for which the standard deviation $\sigma
= 0.11 \div 0.15$ represents the random errors contribution, and the error of
the mean (0.01 $\div$ 0.04) the contribution due to systematic errors. 

\subsection{ Systematic errors}

The relevance of systematic errors is always difficult to reliably assess.
We do not think there are serious concerns related to the adopted $gf$\ scale. On
the other side, uncertainties due to the adopted model atmospheres may be large
since various important aspects (like convection, molecular opacities, and
horizontal inhomogeneities) are far from being adequately known. Large trends
of Fe abundances with excitation have been obtained in the analysis of field
metal-poor giants by Dalle Ore (1992), Dalle Ore et al. (1996), Gratton \&
Sneden (1994), and Gratton et al. (1996), when using both BEGN and K92
model atmospheres. These trends suggest that currently available model
atmospheres are not fully adequate for at least some metal-poor giants
(see e.g. Castelli et al. 1996). While
absolute abundances are quite sensitive to this source of errors, the
comparison of relative abundances obtained with different model atmosphere
grids (K92 and BEGN) suggests that our [Fe/H] values are not heavily affected.
However, our analysis should obviously be repeated once
improved model atmospheres for metal-poor giants become available. 

We need to concern less about possible errors in the adopted temperature scale (in our
case, the CFP one). In fact, were the \teff\ scale largely in error, we would
expect a rather large difference between average abundances provided by neutral
and singly ionized Fe lines. The values listed in Column~2 of
Table~\ref{t:error} indicate that a systematic error of 100~K in the adopted
\teff's would translate into a systematic difference of 0.2~dex between
abundances of Fe~I and Fe~II. Since the observed difference ranges from
0.02~dex to 0.13~dex (depending on the considered sample), we conclude that the
\teff\ scale cannot be systematically incorrect by more than 50~K. 

\subsection{ Internal errors}

Internal errors may be determined from a comparison with the observed scatter in
our abundance determinations (of individual lines and of individual stars in
each cluster). We will consider only errors in the $EW$s and in the adopted
atmospheric parameters, while we regard internal errors in the adopted $gf$s as
negligible.

\subsubsection{ Equivalent widths}

The scatter of abundances from individual (Fe~I) lines is 0.13, 0.11, 0.15, 
0.15, 0.14 and 0.12~dex for the CG96, SKPL, G8689, McW92, M93 and NDC samples 
respectively. These values for the
scatter can be ascribed to errors in the $EW$s of a few m\AA\ (see Section 3), 
and yield mean
internal errors of 0.03 and 0.06~dex for Fe~I and Fe~II respectively. These 
internal errors can be added quadratically and give a prediction
of about 0.07~dex for the scatter in the differences between abundances
derived from Fe~I and Fe~II lines. Since the observed scatter ranges 
from $\sigma =
0.11$\ to $\sigma = 0.15$\ (depending on the adopted sample), additional
sources of errors are clearly present, probably related to the adopted values
for the atmospheric parameters (see Table~\ref{t:error} and discussion below).

\subsection{ Temperatures }

CFP $V-K$\ colours have errors of $\sim 0.05$~mag, which corresponds to
35--40~K using their calibration. This is the internal error of \teff's for
stars within a cluster. When comparing stars in different clusters, the effects
of errors in the interstellar reddening should also be considered. Comparing
various estimates for the same cluster, we estimate an uncertainty of
$0.02\div 0.03$~mag in $E(B-V)$, and 2.7 times larger in $E(V-K)$. Hence,
there is an additional systematic error of $\sim 0.05$~mag in the $(V-K)_0$\
colour ($\sim 35-40$~K) systematic for all stars in a cluster (but random from
cluster to cluster) due to errors in the reddening. If we add these two
uncertainties quadratically, we estimate that the adopted \teff's have internal
errors of $\sim 50$~K. The same figures approximately hold for the $B-V$
colour, which is a less accurate temperature indicator (see $e.g.$, Gratton 
et al. 1996), but at the same time is measured with a precision better by a
factor of 5 than the $V-K$\ for bright globular cluster giants. 

Table~\ref{t:error} suggests that most of the residual scatter in the
differences between Fe~I and Fe~II abundances may be attributed to random
errors in the adopted \teff\ values.

\subsection{ Gravities }

The adopted gravities were deduced from the location of the
stars in the CMD. Since
they were not deduced from the ionization equilibrium, one could think that
errors in $\log g$\ and in \teff\ are not tied\footnote{As a generic estimate,
an error of 100~K in \teff\ translates into a 0.3~dex error in $\log g$, when
deriving spectroscopically the gravities}. But, as matter of fact, temperature
and gravity are not completely independent, since to derive $\log g$\ from the
position of the star in the CMD we have to use the relationship L = $4\pi R^2
\sigma$\teff$^4$, i.e.: 
\begin{equation}
\log g/g_\odot = 4\log T_{\rm eff}/T_\odot - \log L/L_\odot +
\log M/M_\odot 
\end{equation}
To estimate the order of magnitude of the errors affecting gravity, consider
the following:
\begin{itemize}
\item the luminosity $\log L$\ can be wrong either if the cluster distance
moduli or the bolometric corrections $BC$\ are wrong. The distance moduli were
derived by CFP from a compilation of literature data; however uncertainties are
not larger than 0.2 mag, i.e. 0.08~dex in $\log L$. The error in the $BC$s
depends on the model atmospheres, \teff's, and metal abundances. It
cannot be larger than 0.2-0.3 mag, though, 
else the luminosity at the tip of the RGB
would disagree with that predicted by stellar evolution models. This
contribution then translates into another 0.08~dex. If we sum these two
contributions quadratically, we estimate that the total uncertainty in adopted
luminosities is $\sim 0.11$~dex in $\log L$. 
\item the contribution of errors in the mass is very small, since this is fixed
by the age of the globular clusters, and it cannot be wrong by more than 10\%.
The uncertainty in $\log M$\ may then be taken as 0.04~dex. 
\item as mentioned above, uncertainties in \teff's are of $\sim 50$~K, i.e.
$\sim 0.005$~dex in log~\teff.
\end{itemize}
From these considerations, we estimate that the adopted gravities have
internal errors of $\sim 0.15$~dex.

In column 3 of Table~\ref{t:error} we investigate the effects of a variation
of 0.5~dex in the surface gravity; on the basis of the previous discussion, the
contribution from this column should be then divided by at least a factor of 3.
It is interesting to note that a larger error of $\Delta\log g=0.25$\ would
explain the whole residual 0.11~dex in the random error. This is not the case,
though, since there is surely a contribution from errors in \teff: this further
confirms that $\Delta\log g=0.25$\ is an overestimate, and the assumed value of
0.15~dex is reliable. 

\subsection{ Metallicities}

For each star analyzed we have also random errors in the estimate of [A/H] due
to errors in \teff, in gravity (of little entity) and in the measured $EW$s.
This kind of errors can be evaluated from independent analyses of the same
star. To this purpose, we can compare the results obtained for stars in 
the same
cluster, since they are thought to share the same overall metallicity: the
r.m.s deviation from the mean will give an idea of the uncertainties due to
random factors. The quadratic average is 0.06~dex and so they contribute very
little to the observed difference in the abundances from Fe~I and Fe~II (less
than 0.025~dex, from Table~\ref{t:error}). 

\subsection{ Microturbulent velocity}

The internal error in the \vt\ is usually estimated from the comparison of
empirical and theoretical curve-of-growth; it is typically not larger than
0.2~\kms\ for the giants analyzed, since the microturbulent velocity is
derived using Fe~I lines both on the linear and saturation part of the
curve-of-growth. As above, an independent test of the random errors comes from
the comparison between the values obtained for the same star independently
analyzed. We obtained $\Delta$\vt=0.17~\kms\ for the star C428 in
CG96 and in the G8689 sample; it confirms that the microturbulent velocity has
an error smaller than 0.2~\kms. 

\subsection{ Discussion of errors }

To conclude, we have to consider two kinds of errors: first, the internal,
random errors, that affect the comparison from star to star, and second, the
systematic errors, that give an idea of the reliability of our metallicity
scale, of the temperature scale adopted, etc. For the random errors, we have
seen that reasonable estimates are 50~K in \teff, 0.15~dex in $\log g$,
0.06~dex in [A/H] and 0.2~\kms\ in \vt; these errors will affect the
scatter of our data. As to systematic errors, we have only the indication
given by the difference in the abundances from neutral and singly ionized Fe
lines; from the previous discussion, we conclude that these errors are of the
same order of magnitude of random ones. 

Columns 6 and 7 of Table~\ref{t:error} list the uncertainties in the [Fe/H]
ratios derived from the quadratic sum of the contributions from random and
systematic errors, respectively. We remark that the changes in the parameters
used to construct these columns {\it are not} those indicated in the Table, but
the more realistic estimates obtained from the above discussion. From
Table~\ref{t:error} we can estimate that the total uncertainty in our Fe~I
abundances (from which we derive the clusters metallicity) is about 0.11~dex
for the most metal-poor stars, increasing to about 0.13~dex for the most
metal-rich stars. 

\section{The metallicity scale of the globular clusters}

Mean values of the metallicities derived for the 21 clusters analyzed are
listed in Table~\ref{t:clusters}. 

\begin{table*}
\caption{ Mean metallicities for globular clusters compared to literature
data}
\begin{tabular}{llrcccccc}
\hline
\hline 
\\
NGC &Messier &Stars &Mean$\pm$r.m.s. &$\sigma$ &G8689 &SKPL &Others &ZW \\
\\
\hline 
\\
104  &47 Tuc&  5 & -0.70$\pm$0.03 & 0.07 & -0.82 &      &          &-0.71\\
288  &      &  2 & -1.07$\pm$0.03 & 0.04 & -1.31 &      &          &-1.40\\
362  &      &  1 & -1.15          &      & -1.18 &      &          &-1.27\\
1904 &M 79  &  2 & -1.37$\pm$0.00 & 0.01 & -1.42 &      &          &-1.68\\
2298 &      &  3 & -1.74$\pm$0.04 & 0.06 &       &      &-1.91$^a$ &-1.85\\
3201 &      &  3 & -1.23$\pm$0.05 & 0.09 & -1.34 &      &          &-1.61\\
4590 &M 68  &  3 & -1.99$\pm$0.06 & 0.10 & -1.92 &      &-2.17$^b$ &-2.09\\
4833 &      &  3 & -1.58$\pm$0.01 & 0.01 & -1.74 &      &-1.71$^b$ &-1.86\\
5272 &M 3   & 10 & -1.34$\pm$0.02 & 0.06 &       &-1.46 &          &-1.66\\
5897 &      &  2 & -1.59$\pm$0.03 & 0.05 & -1.84 &      &          &-1.68\\
5904 &M 5   & 16 & -1.11$\pm$0.03 & 0.11 & -1.42 &-1.17 &          &-1.40\\
6121 &M 4   &  3 & -1.19$\pm$0.03 & 0.06 & -1.32 &      &          &-1.33\\
6144 &      &  1 & -1.49          &      &       &      &-1.59$^b$ &-1.75\\
6205 &M 13  & 23 & -1.39$\pm$0.01 & 0.07 &       &-1.49 &          &-1.65\\
6254 &M 10  & 15 & -1.41$\pm$0.02 & 0.10 & -1.42 &-1.52 &          &-1.60\\
6341 &M 92  &  9 & -2.16$\pm$0.02 & 0.05 &       &-2.25 &          &-2.24\\
6352 &      &  3 & -0.64$\pm$0.06 & 0.11 & -0.79 &      &          &-0.51\\
6362 &      &  2 & -0.96$\pm$0.00 & 0.01 & -1.04 &      &          &-1.08\\
6397 &      & 10 & -1.82$\pm$0.04 & 0.10 & -1.88 &      &          &-1.91\\
6656 &      &  3 & -1.48$\pm$0.03 & 0.06 & -1.56 &      &          &-1.75\\
6752 &      & 12 & -1.42$\pm$0.02 & 0.08 & -1.53 &      &          &-1.54\\
6838 &M 71  & 13 & -0.70$\pm$0.03 & 0.09 & -0.81 &-0.79 &          &-0.58\\
7078 &M 15  &  4 & -2.12$\pm$0.01 & 0.01 &       &-2.30 &-2.23$^b$ &-2.17\\
7099 &M 30  &  2 & -1.91$\pm$0.00 & 0.00 &       &      &-2.11$^b$ &-2.13\\
\\
\hline
\end{tabular}
\\
References: a = McW92, b = M93
\label{t:clusters}
\end{table*}

The internal uncertainty in [Fe/H] abundances ($\sigma/N^{1/2}$, where $N$\ is
the number of stars studied in each cluster) is very small: on average,
0.06~dex, which can be interpreted also as the mean precision of the cluster
ranking on our new metallicity scale. For comparison, in the same Table, we
also give the original [Fe/H] ratios obtained in previous analyses. In the
last column the metal abundances  from the compilation of Zinn and
West (1984) are listed, superseded and integrated for a few clusters by 
the new measurements 
of Armandroff and Zinn (1988); this scale will be indicated as a whole, 
hereinafter, as ZW. 

\subsection{ Comparison with the ZW scale}

The 24 clusters of Table~\ref{t:clusters} can now be regarded as standard
reference clusters to calibrate individual metal abundance indicators with
metallicities directly derived from high-dispersion spectroscopic analysis. We
feel confident that our list covers fairly well the whole range in metallicity
of globular clusters, going from typical metal-rich clusters, as 47 Tuc, M
71 and NGC 6352, to the classical metal-poor templates (M 92, M 15, M 68). The
sample of intermediate metallicity clusters is also very well represented among
our calibrators. 
One
of our main purposes is to revise and refine the calibration of the ZW ranking
system, which covers almost all known globular
clusters. The main advantage of the ZW scale is that their system is applicable
even to the most distant objects, being based on the integrated parameter
Q$_{39}$ and/or on low-dispersion spectroscopy of the infrared Ca~II triplet.
On the other side, any integrated index is not, by definition, a function of a
single element in a globular cluster. In particular, the major contribution to
line blanketing in the spectral range covered by the $Q_{39}$\ index is due to
the H and K lines of Ca~II, with other significant fractions due to the
$\lambda$~3883 CN band and some Fe blends. Hence, reliability of ZW
metallicities ultimately rests on the coupling between Ca, C, N, and Fe
abundances. It is outside the purposes of this paper to 
proceed further on this point; 
we only wish to recall here that the strength of CN-bands is known to vary from star
to star (the so-called CN-signature), having a bimodal distribution in most
(but not all!) clusters (see Kraft 1994 for a recent review). 

Moreover (see $e.g.$, Clementini et al. 1995, section 5.1.1) the [Ca/Fe] 
ratio does not scale with Fe on the whole range of metallicities, being lower
in metal-rich than in metal-poor Population II stars. Furthermore, a serious
{\it caveat} has been advanced on the claimed independence of the $Q_{39}$\
index from the horizontal branch morphology (see $e.g.$, Smith 1984).
To overcome this kind of problems, the most straightforward way to correct
the ZW scale consists in working directly on the final 
metallicities, since the original compilation of Zinn and West (1984) was
obtained averaging a number of [Fe/H] values derived from different indicators
($e.g.$, $(B-V)_{0,g}$, $\Delta V_{1.4}$, $\Delta S$) and calibrated against 
$Q_{39}$.

\begin{figure}[htb]
\psfig{figure=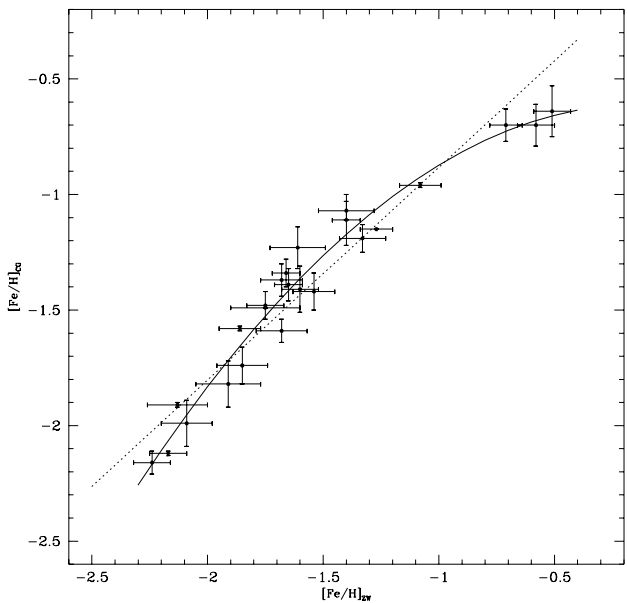,width=8.8cm,clip=}
\caption{ Mean metallicities for the 24 clusters from the present work
compared with metallicities on the Zinn and West scale (1984).}
\label{f:fig5}
\end{figure}

In Figure~\ref{f:fig5} we then compare our high-dispersion [Fe/H] values with
the ZW values for the 24 calibrating clusters. The error bars (1 $\sigma$) are
from Zinn and West (1984: Table 5) and from our Table~\ref{t:clusters}. 
As it is evident from this figure, the ZW scale is far from linear, deviating
both in the low and in the high metallicity regimes, when compared with [Fe/H]
from our direct analysis. In the
metal-rich region ([Fe/H]$>-1$) ZW's metallicities are on average 0.08~dex too
high for the 3 clusters 47 Tuc, M 71 and NGC 6352, with the last two objects
being responsible for most of the discrepancy (0.12 and 0.13~dex,
respectively). For $-1 \le $[Fe/H]$ \le -1.9$ the [Fe/H] values of ZW are
definitively too low by a mean value of 0.23~dex ($\sigma$=0.09~dex, 16
clusters). Finally, in the very low-metallicity tail, ZW's values are on
average 0.11~dex higher than ours ($\sigma$=0.06~dex, 6 clusters). 

The non-linear behaviour has
been confirmed by a $t-$test on the significance of the quadratic term in the
relation between ZW and ours [Fe/H]'s. 
To bring ZW's [Fe/H] ratios on a metallicity scale fully based only on high
dispersion spectroscopy (HDS) we then derived a correction given by a quadratic
relation. This procedure automatically takes into account also the different
zero point between the two scales, since the ZW scale was ultimately based on
the Cohen (1983) scale, which, as other past analysis, adopts the 
traditional old solar Fe abundance log $\epsilon$=7.67.
The resulting function we derive for this correction is: 
\begin{eqnarray}
{\rm [Fe/H]_{CG}} & = - 0.618 (\pm 0.083) \nonumber \\
    & - 0.097\,(\pm 0.189){\rm [Fe/H]_{ZW}} \nonumber \\
    & -0.352\,(\pm 0.067){\rm [Fe/H]_{ZW}}^2
\end{eqnarray}
with the correlation coefficient $r=0.982$\ and $\sigma$=0.08 for 24 clusters.
This relationship is highly significant, from a statistical point of view, and
can be applied to ZW metallicities in the range $-2.24<$[Fe/H]$_{\rm ZW}<-0.51$,
defined by the lowest and highest values of [Fe/H]$_{\rm ZW}$\ among the
clusters used for the calibration. The quadratic regression line is shown as a
heavy line in Figure~\ref{f:fig5}; overimposed in the same Figure is also the
result of a linear fit, which takes into account the errors. As one can see,
even considering 3$\sigma$\ error bars, it it very difficult to represent the
data on a linear scale, in particular at the lower metallicity edge. 

Once the correction is applied the non-linearity of the ZW scale obviously
disappears. However, a certain
amount of scatter is still present in the intermediate-metallicity regime; we
believe that it could be attributed to a residual effect, not well removed by
our calibration, of the second parameter. This last is in fact likely to affect
ZW's metallicities more severely in this regime, in which the integrated
colours of clusters of different HB morphological type can be sensibly
misinterpreted in terms of [Fe/H].

The next logical step would be now to calibrate other empirical metallicity
indicators, $i.e.$ repeat the original work of Zinn and West (1984) but 
using now our direct [Fe/H] values from HDS as a calibrating sequence. 
The most interesting and accessible parameters are the photometric ones ($e.g.$,
(B-V)$_{0,g}$, $\Delta V_{1.4}$, etc.): they are widely used since it is easy
enough to measure them from the recent and accurate CCD-based colour-magnitude
diagrams (CMD). However, it would be preferable
to have a dataset of homogeneity and accuracy comparable
with the precision of our metallicities, instead of relying on compilation 
from different sources.
Since such an effort is presently in progress on a
set of CMDs analyzed in a self-consistent way, we postpone to a forthcoming
paper this kind of calibration.
However, an immediate and meaningful comparison can be made with 
the metallicity scale for globular clusters derived from RR Lyraes, since
we can compare results obtained for two different stellar populations, RGB
stars and HB stars, independently checking the validity of both scales.

\subsection{Comparison with the metallicity scale of RR Lyrae stars}

\begin{figure}[htbp]
\psfig{figure=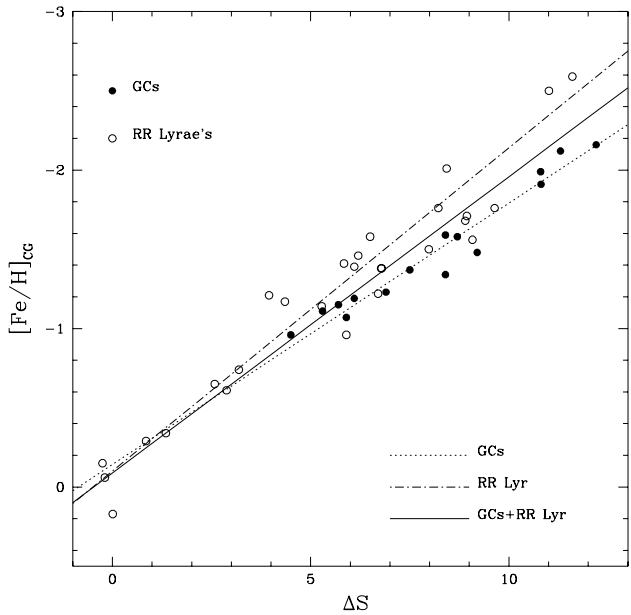,width=8.8cm,clip=}
\caption{ Calibration of the $\Delta S$\ index with our new analysis and 
with the data of field RR Lyrae variables from Clementini et al. (1995)}
\label{f:fig7}
\end{figure}

The most recent calibration of [Fe/H] in terms of the Preston's (1959) 
index $\Delta S$\ is the one
defined by Clementini et al. (1995), who found
\begin{equation}
{\rm [Fe/H] = -0.194 (\pm 0.011) \Delta S -0.08 (\pm 0.18) }
\end{equation}
This relation was derived using RR Lyraes both in the field and in globular
clusters.
However, while metallicity values for field RR Lyraes were directly derived
from high-resolution spectra or from the re-analysis of literature data (for a 
total of 28 RR Lyraes), cluster
metallicities were taken at face value from the literature, even if a zero
point was admittedly noted while using data from different samples. We 
have many clusters in
common with the study of Clementini et al. (1995) and have then derived 
again the
[Fe/H] $vs$ $\Delta S$\ relation. Figure~\ref{f:fig7} shows the result of our
re-analysis.

We obtained $\Delta S$ values for 15 of our calibrating clusters from the
metallicities of Costar \& Smith (1988), inverting the Butler's (1975) relation
they used. Our values are not completely identical to those used by Clementini
et al. (1995); the main differences are that a) we excluded 47 Tuc, since its
value for $\Delta S$\ is based on a single star, possibly not member of the
cluster (Tucholke 1992) and b) for NGC 288 we assumed a value of $\Delta S=5.9$,
since the mean value 7.2 cited by Costar \& Smith was obtained including
spectra taken at phases near maximum light. 

Regression lines were then obtained by least-squares fits (we averaged values
obtained exchanging the independent and dependent variables):
\begin{itemize}
\item If we consider only the 15 GCs we obtain:
\begin{equation}
{\rm [Fe/H]}=-0.165\,(\pm ~0.019)\,\Delta S-0.142\,(\pm 0.033)
\end{equation}
with a $\sigma=0.130$, and $r=0.947$\ (dotted line in Figure~\ref{f:fig7}).
\item It we consider both the 15 GCs and the 28 field RR Lyraes we get:
\begin{equation}
{\rm [Fe/H]}=-0.187\,(\pm 0.011)\,\Delta S-0.088\,(\pm 0.041)
\end{equation}
with a $\sigma=0.269$, and $r=0.954$\ (solid line in Figure~\ref{f:fig7}).
\end{itemize}

Also shown in Figure~\ref{f:fig7} is the calibration obtained by Clementini et
al. (1995), using only 28 field variables (their equation 6): 
[Fe/H]$=-0.204$($\pm$0.012)$\Delta$S $-0.102$($\pm$0.036), $\sigma$=0.190. 

The first striking evidence both from Figure~\ref{f:fig7} and the above
equations is that the sequence of the globular cluster points seems to be much
better defined, with a smaller scatter than the distribution of field RR
Lyraes. The scatter in equation (6) of Clementini et al. 
is 0.19~dex, to be compared with the value of
0.13~dex obtained using only the new values for the clusters. We stress the
fact that both the solar Fe abundance and the source for the oscillator
strengths are in common between the present analysis and that of Clementini et
al. (1995); moreover, the procedure followed in the abundance analysis is
virtually the same. This may be evidence in favour of a larger intrinsic scatter
in field than in cluster variables, or it may just reflect a smaller
error in the values of $\Delta S$\ for cluster RR Lyraes. However the last
explanation seems a little unpalatable, since determinations of $\Delta S$\
values are usually more accurate for nearby field stars. 

The second feature shown in Figure~\ref{f:fig7} is a rather clear separation between the
relations for cluster and field RR Lyraes in the low metallicity region; this is
the likely explanation for the increase in the scatter when the calibration
[Fe/H]--$\Delta$S is made using both cluster and field variables.
The same behaviour was evident also in Figure 14c of Clementini et al. (1995),
but here it is even clearer, given the high degree of homogeneity in
our data. Why this is so, we are not sure, apart from 
a suggestion of non-linearity
in the $\Delta S$-[Fe/H] relation theoretically predicted (Manduca 1981)
and discussed in Clementini et al. (1995).
Apart from this, there seems to be a good 
agreement between both scales; if we use our new calibration (equation 12) to
derive [Fe/H] ratios, the differences [Fe/H]$_{\rm CG}$ $-$ [Fe/H]$_{\Delta S}$
are on average $0.12\pm 0.03$ ($\sigma$=0.10, for 16 clusters).

\section{Discussion and conclusions}

We think that the effort to get more reliable and accurate metallicities 
is truly
worthwhile: a new, homogeneous scale of [Fe/H] for globular
clusters is really needed, since up to now the numerous but still 
heterogeneous estimates from HDS have been systematically ignored in many 
problems of stellar evolution.

\begin{figure}[htbp]
\psfig{figure=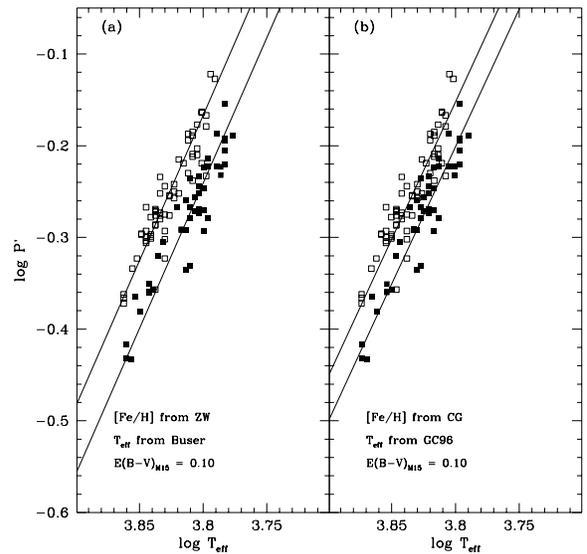,width=8.8cm,clip=}
\caption{ Observed distributions of variables in M 3 (filled squares) and 
M 15 (open squares) in the log P'--
log T$_{\rm eff}$ plane (see text).}
\label{f:fig8}
\end{figure}

As an illustrative example 
we will explore the effects of our new [Fe/H] values on the long-standing
problem of the Oosterhoff (1944) effect among globular clusters, which
belong to 2 groups on the basis of the mean period of their type $ab$ RR Lyraes;
this division reflects a separation in metallicity of the clusters (Arp 1955).
The statement of this problem, its history and references are fully addressed 
in a recent paper by Sandage (1993a). Briefly, the concept is to use
the pulsation equation for RR Lyrae stars (P = f (T$_{\rm eff}$,L,M), 
where P is the fundamental period of the pulsation, L the luminosity and M the mass of the
star) to derive a calibration of the absolute luminosity of RR Lyraes in terms
of the metal content, [Fe/H], in turn a cornerstone to ultimately get 
the ages of globular clusters. This is possible if the
variations of the parameters involved (P,T$_{\rm eff}$, L, M) 
with [Fe/H] are known.

The critical and more debated point is to determine how $\Delta$logP, the
shift existing in the log P-log \teff\ plane between the distribution of 
variables in OoI and OoII clusters, varies with the metallicity; the slope, in
particular, is still much controversial. While Sandage (1993a) finds a value
of about 0.12, theoretical models seem to predict a much lower value, 
around 0.05 (in the sense of longer periods for metal-poorer clusters).
Our approach is based both on our new metallicities and new \teff's 
for variables in
M 3 (OoI) and M 15 (OoII), $i.e.$ the template-pair for the Oosterhoff effect.
\footnote{ Our conclusions would not change had we used M 68
([Fe/H]$=-1.99$ on our scale) instead of 
M 15 as OoII template. While M 68 has a small and well determined
reddening ($E(B-V)$=0.03), M 15 has a better populated instability strip.}

(a) From the present work we adopt [Fe/H]$=-1.34$ 
for M 3 and [Fe/H]$=-2.12$ for M 15.  These values compare well
to those derived by Sneden et al. (1991) and Kraft et al.
(1992, [Fe/H]$=-1.48$ and $-$2.30 for M 3 and M 15 respectively), allowing for 
their use of the old
Bell et al. (1976) model atmospheres and of a different set of $gf$.

If one compares these new values with the classical ZW ones ($-$1.66 for M 3 and 
$-$2.15 for M 15) ,
it it immediately evident that
{\it whatever the period shift between M 3 and M 15 may be, 
it has to be ``diluted''
on a larger range of metallicity than before}. This in turn affects the
slope of the relationship log P--[Fe/H].

b) We adopt here the new temperature scale derived by Gratton et al. (1996).
Briefly, they first derived empirical
colour-\teff\ relations for population~I stars, based on
\teff's from the Infrared Flux Method (IRFM, Blackwell and Lynas Gray 1994; 
Bell and Gustafsson 1989)
and interferometric diameters (Di Benedetto and Rabbia 1987); stars of 
luminosity class III and V were considered separately.
T$_{\rm eff}$'s for stars of different log $g$ and [Fe/H] are then obtained
applying systematic corrections to the theoretical K92 T$_{\rm eff}$'s, 
to tie them to the empirical calibrations and to the K92 colours, in
order to take into account the real metallicities (different from 
solar). The effect of gravity was taken into account by interpolating
between typical values for dwarfs and giants.
The underlying assumption is that K92 models (the same consistently used in
the abundance analysis) are well able to
reproduce the run of colours with the overall metal content [A/H], but have to
be corrected in function of the surface gravity
(a constant mean value of log $g=2.75$ was assumed for all variables).

The adopted scale is very similar to the one defined by Clementini et al. (1995)
to study field RR Lyrae stars; they also showed that as far as colours (and 
abundances) are concerned, K92 models are well suited to reproduce the 
atmospheres of RR Lyrae variables.\footnote{The reader should be aware of the
small inconsistency due to our use of the CFP scale for globular cluster 
(cooler) giants and the new temperature scale for RR Lyraes; however, as 
discussed in Gratton et al. (1996), differences between the two scales are well
within the intrinsic uncertainties of both scales and do not alter
significatively our conclusions.}

The photometry for the variables of M 15 is taken from the high quality work
of Bingham et al. (1984). For M 3, we used colours from Sandage (1990).
The choice of the mean colour which better represents 
the one the variable should have were it not pulsating is not simple (see 
$e.g.$, Fernley 1994). However, for sake of comparison with Sandage's
previous works on M 3, in the present study we use $(<B>-<V>)_{\rm corr}$,
the corrected colours taken 
from the quoted sources; they also include an empirical correction $\Delta$C(A) 
depending on the light-curve  amplitude (Sandage 1990).

We adopt $E(B-V)$=0.00 for M 3, while for M 15 we assume 0.10 (Zinn 1980).

Figure~\ref{f:fig8} shows the observed distributions of variables in M 3 
and M 15 in the
classical log P'--log T$_{\rm eff}$ plane (log P' is the reduced
and fundamentalized period; for c-type RR Lyraes the period is fundamentalized
by adding 0.127 to log P). The lines are best-fits drawn by eye 
through the data, since the computed least squares linear regression result
in non-parallel lines, due to the scatter among the data, especially in 
those of M 3. This method is good enough in this case, since we are only
interested in presenting and emphasizing another source of uncertainty
affecting this method. 

In panel (a) T$_{\rm eff}$'s are derived from the old Kurucz (1979) model
atmospheres, following the (unpublished) calibration of Buser, and with 
$E(B-V)$=0.00 and 0.10 for M 3 and M 15 respectively. 
Metallicities are on 
the Zinn and  West's scale: this panel should then reproduce
closely enough Sandage's results for the Oosterhoff effect in the pair
M 3-M 15. In panel (b) we used the new metallicities from the present work 
and T$_{\rm eff}$'s derived from K92 models. 

In order to evaluate the period shift between the two distributions, we 
followed the prescriptions of Sandage (1993b), reading the values of log P'
not at constant log T$_{\rm eff}$, but at lower temperatures for lower 
metallicities, following his equation (5).
Our measurements then give the period shift $\Delta$logP (in the sense 
M 15-M 3) = $-$0.103 and $-$0.076 for case (a) and (b), respectively.
We note that only the first slope is similar to the one 
derived by Sandage (1993a: $\Delta$logP/$\Delta$ [Fe/H]$=-0.122$). 
Simply using the new metallicities
and effective temperatures, although derived from the same photometric data,
significatively decreases the size of the effect. 

Taking at face value these figures, and using the parameters temperature
and mass (along with their variations with metallicity as given by Sandage 
1993b) in the pulsation equation of van Albada and Baker (1973), we obtain
for the slope of the relationship M$_{\rm bol}$ $vs$ [Fe/H] the values 
0.300 and 0.221 for case (a) and (b) respectively.
As one can see, case (b) is very similar to that derived from the 
Baade-Wesselink analyses (0.25, as quoted for instance in Sandage 1993b), whose
results were considered up to now in serious disagreement with the calibration
from the Oosterhoff effect.

This straightforward exercise 
points out that 
simply using new, modern, and
self-consistent temperatures and metallicities, the size of the 
Oosterhoff effect in the template pair M 3-M 15 is somewhat reduced, being
more consistent with theoretical models of Zero Age Horizontal Branch (see
Sandage 1993b for references).
We then conclude that our effort in obtaining these new, improved values for
metallicities can pay off; we will proceed to a new calibration of photometric /
low resolution indices, including all known globular clusters, as soon as a 
homogeneous dataset will be available.

\begin{acknowledgements}
We wish to thank Chris Sneden for useful discussion and encouragement. E.C.
warmly thanks Carla Cacciari for providing tables with the parameters of
variables in M 3 and M 15 from Kurucz's (1979) models and for many useful 
suggestions and discussions on the Oosterhoff effect. E.C. acknowledges 
financial support from the {\it Consiglio Nazionale delle Ricerche}.

\end{acknowledgements}

\begin{table*}
\begin{center}
\caption{ Adopted atmospheric parameters and results of abundance analysis 
for globular cluster giants}
{\tiny
\begin{tabular}{rrrrrrrrrrrrrr}
\hline\hline
\\
NGC  & Star     & Source& Teff & log g & [A/H] &  Vt  &   r & FeI  &$\sigma$(FeI)&   r & FeII &$\sigma$(FeII)&[Fe/H] \\
\\
\hline
\\
104  & 1407     & G8689 & 4500 & 2.00  & $-$0.73 & 1.53 &  18 & 6.79 & 0.12 &   1 & 6.84 &       & $-$0.73 \\
104  & 8416     & G8689 & 4425 & 1.80  & $-$0.82 & 1.79 &  12 & 6.70 & 0.17 &   2 & 6.66 & 0.28  & $-$0.82 \\
104  & 8406     & CG95  & 4040 & 1.10  & $-$0.61 & 1.76 &  35 & 6.92 & 0.18 &   3 & 6.70 & 0.10  & $-$0.60 \\
104  & 5422     & CG95  & 4090 & 1.20  & $-$0.70 & 1.62 &  35 & 6.83 & 0.15 &   2 & 6.90 & 0.01  & $-$0.69 \\
104  & 5529     & CG95  & 3850 & 0.70  & $-$0.66 & 1.59 &   6 & 6.86 & 0.16 &   2 & 6.88 & 0.26  & $-$0.66 \\
288  & 245      & G8689 & 4400 & 1.20  & $-$1.10 & 1.55 &  21 & 6.41 & 0.15 &     &      &       & $-$1.11 \\
288  & 231      & G8689 & 4475 & 1.30  & $-$1.03 & 1.42 &  19 & 6.49 & 0.19 &     &      &       & $-$1.03 \\
362  & I-23     & G8689 & 4335 & 1.10  & $-$1.15 & 1.55 &  23 & 6.37 & 0.16 &     &      &       & $-$1.15 \\
1904 & 223      & G8689 & 4250 & 0.75  & $-$1.36 & 1.69 &  20 & 6.16 & 0.19 &   7 & 6.16 & 0.21  & $-$1.36 \\
1904 & 153      & G8689 & 4270 & 0.75  & $-$1.38 & 2.23 &  17 & 6.15 & 0.15 &     &      &       & $-$1.37 \\
2298 & 7        & MCW92 & 4351 & 0.80  & $-$1.64 & 1.49 &  24 & 5.89 & 0.17 &   2 & 5.77 & 0.14  & $-$1.63 \\
2298 & 8        & MCW92 & 4358 & 0.80  & $-$1.79 & 1.48 &  23 & 5.74 & 0.13 &   2 & 5.51 & 0.12  & $-$1.78 \\
2298 & 9        & MCW92 & 4348 & 1.00  & $-$1.81 & 1.51 &  23 & 5.70 & 0.15 &   2 & 5.65 & 0.07  & $-$1.82 \\
3201 & 3522     & G8689 & 4450 & 1.20  & $-$1.35 & 1.76 &  17 & 6.17 & 0.10 &   4 & 6.22 & 0.02  & $-$1.35 \\
3201 & 1501     & G8689 & 4445 & 1.20  & $-$1.15 & 1.58 &  18 & 6.36 & 0.15 &   5 & 6.34 & 0.21  & $-$1.16 \\
3201 & 1410     & G8689 & 4480 & 1.20  & $-$1.19 & 0.98 &  11 & 6.33 & 0.19 &     &      &       & $-$1.19 \\
4590 & 260      & M93   & 4329 & 0.70  & $-$2.09 & 2.16 &  27 & 5.42 & 0.11 &   3 & 5.23 & 0.15  & $-$2.10 \\
4590 & 53       & M93   & 4400 & 1.00  & $-$2.09 & 1.79 &  26 & 5.42 & 0.13 &   3 & 5.14 & 0.21  & $-$2.10 \\
4590 & I-260    & G8689 & 4329 & 0.75  & $-$1.63 & 0.82 &  11 & 5.90 & 0.13 &   2 & 5.88 & 0.01  & $-$1.63 \\
4590 & I-256    & G8689 & 4438 & 0.80  & $-$2.01 & 1.66 &   7 & 5.51 & 0.05 &   3 & 5.70 & 0.06  & $-$2.01 \\
4833 & 13       & M93   & 4500 & 1.30  & $-$1.59 & 1.78 &  23 & 5.93 & 0.12 &   1 & 5.74 &       & $-$1.59 \\
4833 & B172     & G8689 & 4323 & 0.80  & $-$1.56 & 1.72 &  18 & 5.96 & 0.16 &   5 & 6.08 & 0.13  & $-$1.56 \\
4833 & MA-1     & G8689 & 4273 & 0.75  & $-$1.59 & 1.90 &  21 & 5.94 & 0.14 &   6 & 6.10 & 0.19  & $-$1.58 \\
5272 & II-46    & SKPL  & 4000 & 0.60  & $-$1.41 & 2.02 &  15 & 6.12 & 0.06 &   4 & 6.14 & 0.14  & $-$1.40 \\
5272 & 297      & SKPL  & 4070 & 0.70  & $-$1.39 & 1.93 &  12 & 6.14 & 0.16 &   5 & 6.14 & 0.10  & $-$1.38 \\
5272 & AA       & SKPL  & 4000 & 0.40  & $-$1.32 & 2.09 &  17 & 6.20 & 0.11 &   5 & 6.23 & 0.07  & $-$1.32 \\
5272 & MB-3     & SKPL  & 3900 & 0.20  & $-$1.30 & 1.76 &  14 & 6.22 & 0.15 &   4 & 6.38 & 0.11  & $-$1.30 \\
5272 & MB-4     & SKPL  & 3925 & 0.30  & $-$1.23 & 1.93 &  18 & 6.30 & 0.20 &   4 & 6.39 & 0.21  & $-$1.22 \\
5272 & 1000     & SKPL  & 4175 & 0.45  & $-$1.38 & 1.98 &  15 & 6.14 & 0.11 &   4 & 6.20 & 0.14  & $-$1.38 \\
5272 & 1127     & SKPL  & 4225 & 0.90  & $-$1.34 & 1.80 &  17 & 6.19 & 0.11 &   4 & 6.14 & 0.10  & $-$1.33 \\
5272 & 1397     & SKPL  & 3950 & 0.40  & $-$1.32 & 2.13 &  17 & 6.20 & 0.12 &   4 & 6.20 & 0.12  & $-$1.32 \\
5272 & III-28   & SKPL  & 4160 & 0.75  & $-$1.45 & 1.69 &  14 & 6.07 & 0.15 &   5 & 6.10 & 0.18  & $-$1.45 \\
5272 & MB-1     & SKPL  & 3825 & 0.00  & $-$1.34 & 2.34 &   8 & 6.19 & 0.18 &   2 & 6.24 & 0.08  & $-$1.33 \\
5897 & 9        & G8689 & 4175 & 0.75  & $-$1.64 & 1.86 &  12 & 5.88 & 0.13 &   2 & 6.08 & 0.03  & $-$1.64 \\
5897 & 160      & G8689 & 4300 & 0.80  & $-$1.55 & 1.67 &  17 & 5.98 & 0.13 &   3 & 6.17 & 0.17  & $-$1.55 \\
5904 & IV-47    & SKPL  & 4110 & 0.90  & $-$1.04 & 1.75 &  18 & 6.48 & 0.11 &   5 & 6.48 & 0.11  & $-$1.04 \\
5904 & II-85    & SKPL  & 4050 & 0.90  & $-$1.00 & 1.75 &  19 & 6.52 & 0.11 &   5 & 6.54 & 0.07  & $-$1.00 \\
5904 & III-122  & SKPL  & 4050 & 0.70  & $-$1.09 & 2.05 &  16 & 6.43 & 0.12 &   4 & 6.46 & 0.15  & $-$1.09 \\
5904 & III-3    & SKPL  & 4070 & 0.75  & $-$1.15 & 1.99 &  18 & 6.37 & 0.10 &   4 & 6.46 & 0.12  & $-$1.15 \\
5904 & III-96   & SKPL  & 4300 & 1.30  & $-$1.07 & 1.61 &  18 & 6.46 & 0.09 &   5 & 6.41 & 0.07  & $-$1.06 \\
5904 & IV-72    & SKPL  & 4300 & 1.20  & $-$1.01 & 1.59 &  16 & 6.51 & 0.10 &   5 & 6.45 & 0.06  & $-$1.01 \\
5904 & II-9     & SKPL  & 4300 & 0.80  & $-$0.98 & 1.78 &  13 & 6.54 & 0.16 &   3 & 6.64 & 0.08  & $-$0.98 \\
5904 & I-68     & SKPL  & 4130 & 0.90  & $-$1.09 & 2.01 &  17 & 6.44 & 0.15 &   4 & 6.43 & 0.15  & $-$1.09 \\
5904 & III-78   & SKPL  & 4200 & 1.00  & $-$1.03 & 1.86 &  18 & 6.49 & 0.16 &   6 & 6.46 & 0.09  & $-$1.03 \\
5904 & III-36   & SKPL  & 4250 & 1.10  & $-$1.06 & 1.61 &  18 & 6.46 & 0.11 &   5 & 6.40 & 0.08  & $-$1.06 \\
5904 & I-4      & G8689 & 4385 & 1.30  & $-$1.41 & 1.97 &  21 & 6.11 & 0.10 &   3 & 6.26 & 0.04  & $-$1.41 \\
5904 & I-25     & G8689 & 4435 & 1.40  & $-$1.14 & 1.68 &  13 & 6.37 & 0.16 &     &      &       & $-$1.15 \\
5904 & I-61     & G8689 & 4380 & 1.30  & $-$1.31 & 1.92 &  23 & 6.21 & 0.12 &   4 & 6.45 & 0.07  & $-$1.31 \\
5904 & IV-59    & SKPL  & 4320 & 1.10  & $-$1.12 & 2.19 &  15 & 6.39 & 0.12 &   3 & 6.36 & 0.06  & $-$1.13 \\
5904 & I-71     & SKPL  & 4390 & 1.30  & $-$1.04 & 1.55 &  17 & 6.48 & 0.07 &   5 & 6.44 & 0.11  & $-$1.05 \\
5904 & III-50   & SKPL  & 4400 & 1.00  & $-$1.22 & 1.96 &  18 & 6.30 & 0.11 &   5 & 6.30 & 0.11  & $-$1.22 \\
6121 & 2626     & G8689 & 4295 & 1.60  & $-$1.27 & 1.50 &  20 & 6.26 & 0.16 &   2 & 6.65 & 0.01  & $-$1.26 \\
6121 & 2608     & G8689 & 4285 & 1.50  & $-$1.14 & 1.30 &  21 & 6.38 & 0.10 &   2 & 6.66 & 0.08  & $-$1.14 \\
6121 & 1605     & G8689 & 4270 & 1.50  & $-$1.17 & 1.32 &  22 & 6.36 & 0.14 &   1 & 6.40 &       & $-$1.16 \\
6144 & 152      & M93   & 4600 & 1.50  & $-$1.49 & 1.53 &  20 & 6.03 & 0.13 &   2 & 6.04 & 0.12  & $-$1.49 \\
6205 & 262      & SKPL  & 4180 & 0.80  & $-$1.40 & 1.90 &  18 & 6.13 & 0.10 &   5 & 6.12 & 0.11  & $-$1.39 \\
6205 & III-56   & SKPL  & 4100 & 0.65  & $-$1.41 & 2.01 &  17 & 6.11 & 0.10 &   5 & 6.16 & 0.10  & $-$1.41 \\
6205 & 853      & SKPL  & 4180 & 0.80  & $-$1.39 & 1.82 &  17 & 6.14 & 0.08 &   5 & 6.08 & 0.09  & $-$1.39 \\
6205 & 598      & SKPL  & 3900 & 0.00  & $-$1.45 & 2.11 &  17 & 6.08 & 0.15 &   4 & 6.17 & 0.12  & $-$1.44 \\
6205 & 261      & SKPL  & 4230 & 0.85  & $-$1.36 & 2.01 &  17 & 6.16 & 0.09 &   4 & 6.03 & 0.05  & $-$1.36 \\
6205 & 835      & SKPL  & 4090 & 0.70  & $-$1.23 & 1.82 &  14 & 6.29 & 0.15 &   4 & 6.28 & 0.08  & $-$1.23 \\
6205 & II-90    & SKPL  & 4000 & 0.60  & $-$1.42 & 2.13 &  10 & 6.11 & 0.09 &   5 & 6.16 & 0.14  & $-$1.41 \\
6205 & II-67    & SKPL  & 3950 & 0.30  & $-$1.47 & 2.20 &  17 & 6.05 & 0.14 &   5 & 6.08 & 0.10  & $-$1.47 \\
6205 & IV-25    & SKPL  & 4000 & 0.30  & $-$1.39 & 2.24 &  13 & 6.14 & 0.07 &   4 & 6.12 & 0.08  & $-$1.38 \\
6205 & 940      & SKPL  & 4070 & 0.65  & $-$1.41 & 2.09 &  15 & 6.12 & 0.09 &   4 & 6.24 & 0.06  & $-$1.40 \\
6205 & 324      & SKPL  & 4050 & 0.50  & $-$1.48 & 2.47 &  18 & 6.04 & 0.08 &   4 & 6.20 & 0.15  & $-$1.48 \\
6205 & I-48     & SKPL  & 3920 & 0.30  & $-$1.31 & 2.01 &  16 & 6.21 & 0.10 &   2 & 6.13 & 0.16  & $-$1.31 \\
6205 & III-59   & SKPL  & 4360 & 1.10  & $-$1.33 & 1.68 &  18 & 6.19 & 0.10 &   5 & 6.16 & 0.16  & $-$1.33 \\
6205 & III-52   & SKPL  & 4335 & 1.00  & $-$1.42 & 1.92 &  17 & 6.10 & 0.14 &   5 & 6.13 & 0.12  & $-$1.42 \\
6205 & III-63   & SKPL  & 4200 & 0.80  & $-$1.49 & 2.11 &  18 & 6.02 & 0.09 &   5 & 6.09 & 0.08  & $-$1.50 \\
6205 & L629     & SKPL  & 3950 & 0.20  & $-$1.49 & 2.3  &  15 & 6.03 & 0.08 &   4 & 6.08 & 0.09  & $-$1.49 \\
6205 & III-73   & SKPL  & 4300 & 0.85  & $-$1.34 & 2.10 &  16 & 6.18 & 0.11 &   3 & 6.09 & 0.10  & $-$1.34 \\
6205 & III-18   & SKPL  & 4350 & 1.20  & $-$1.35 & 1.77 &  16 & 6.16 & 0.13 &   4 & 6.13 & 0.25  & $-$1.36 \\
6205 & II-33    & SKPL  & 4360 & 1.15  & $-$1.37 & 1.90 &  18 & 6.15 & 0.11 &   4 & 6.08 & 0.15  & $-$1.37 \\
6205 & I-13     & SKPL  & 4290 & 1.00  & $-$1.32 & 1.81 &  18 & 6.20 & 0.09 &   5 & 6.20 & 0.08  & $-$1.32 \\
6205 & II-34    & SKPL  & 4190 & 0.85  & $-$1.36 & 1.71 &  18 & 6.16 & 0.11 &   5 & 6.10 & 0.14  & $-$1.36 \\
6205 & II-76    & SKPL  & 4350 & 1.00  & $-$1.49 & 2.09 &  18 & 6.04 & 0.08 &   5 & 6.07 & 0.11  & $-$1.48 \\
6205 & II-57    & SKPL  & 4410 & 1.20  & $-$1.38 & 1.64 &  18 & 6.14 & 0.08 &   5 & 6.09 & 0.15  & $-$1.38 \\
6254 & G        & SKPL  & 4650 & 1.20  & $-$1.42 & 1.79 &  19 & 6.09 & 0.16 &   4 & 5.99 & 0.07  & $-$1.43 \\
6254 & II-217   & G8689 & 4410 & 1.20  & $-$1.42 & 1.86 &  26 & 6.10 & 0.15 &   5 & 6.27 & 0.12  & $-$1.42 \\
\\
\hline
\end{tabular}
\label{t:atmos}
}
\end{center}
\end{table*}

\begin{table*}
\begin{center}
{\tiny
\centerline{(cont.)}
\begin{tabular}{rrrrrrrrrrrrrr}
\hline\hline
\\
NGC  & Star     & Source& Teff & log g & [A/H] &  Vt  &   r & FeI  &$\sigma$(FeI)&   r & FeII &$\sigma$(FeII)&[Fe/H] \\
\\
\hline
\\
6254 & I-15     & G8689 & 4405 & 1.00  & $-$1.18 & 1.45 &  24 & 6.33 & 0.13 &   2 & 6.36 & 0.02  & $-$1.19 \\
6254 & H-I-367  & SKPL  & 4135 & 0.60  & $-$1.55 & 1.85 &  18 & 5.98 & 0.10 &   5 & 6.01 & 0.09  & $-$1.54 \\
6254 & A-III-16 & SKPL  & 4150 & 0.90  & $-$1.38 & 1.74 &  17 & 6.15 & 0.11 &   5 & 6.07 & 0.12  & $-$1.37 \\
6254 & A-II-24  & SKPL  & 4050 & 0.10  & $-$1.41 & 1.96 &  16 & 6.11 & 0.10 &   5 & 6.03 & 0.11  & $-$1.41 \\
6254 & A-I-61   & SKPL  & 4550 & 1.00  & $-$1.57 & 2.25 &  15 & 5.95 & 0.09 &   5 & 5.85 & 0.10  & $-$1.57 \\
6254 & H-I-15   & SKPL  & 4225 & 0.75  & $-$1.44 & 1.72 &  18 & 6.08 & 0.11 &   5 & 6.04 & 0.10  & $-$1.44 \\
6254 & A-III-5  & SKPL  & 4400 & 1.20  & $-$1.19 & 1.63 &  18 & 6.32 & 0.10 &   4 & 6.26 & 0.12  & $-$1.20 \\
6254 & A-I-60   & SKPL  & 4400 & 1.10  & $-$1.41 & 1.55 &  19 & 6.11 & 0.09 &   5 & 6.06 & 0.17  & $-$1.41 \\
6254 & A-III-21 & SKPL  & 4060 & 0.50  & $-$1.39 & 2.00 &  17 & 6.13 & 0.12 &   5 & 6.06 & 0.10  & $-$1.39 \\
6254 & D        & SKPL  & 4200 & 1.05  & $-$1.32 & 1.78 &  16 & 6.21 & 0.11 &   5 & 6.11 & 0.12  & $-$1.31 \\
6254 & C        & SKPL  & 4200 & 0.75  & $-$1.54 & 1.84 &  18 & 5.99 & 0.07 &   5 & 5.94 & 0.05  & $-$1.53 \\
6254 & E        & SKPL  & 4350 & 0.80  & $-$1.52 & 1.99 &  18 & 6.01 & 0.10 &   5 & 5.93 & 0.03  & $-$1.51 \\
6254 & B        & SKPL  & 4150 & 0.50  & $-$1.40 & 1.81 &  18 & 6.12 & 0.07 &   5 & 6.03 & 0.20  & $-$1.40 \\
6254 & A-I-2    & SKPL  & 3975 & 0.00  & $-$1.39 & 2.05 &  18 & 6.14 & 0.09 &   5 & 6.11 & 0.14  & $-$1.39 \\
6341 & XI-19    & SKPL  & 4525 & 1.20  & $-$2.10 & 1.83 &   9 & 5.43 & 0.12 &   3 & 5.47 & 0.07  & $-$2.09 \\
6341 & III-82   & SKPL  & 4600 & 1.47  & $-$2.09 & 1.75 &   7 & 5.43 & 0.13 &   2 & 5.50 & 0.22  & $-$2.09 \\
6341 & VII-122  & SKPL  & 4350 & 0.85  & $-$2.15 & 1.95 &   8 & 5.36 & 0.07 &   3 & 5.27 & 0.08  & $-$2.16 \\
6341 & VII-18   & SKPL  & 4230 & 0.70  & $-$2.22 & 1.99 &  10 & 5.31 & 0.09 &   5 & 5.35 & 0.07  & $-$2.21 \\
6341 & III-13   & SKPL  & 4125 & 0.75  & $-$2.17 & 1.98 &  10 & 5.36 & 0.12 &   4 & 5.44 & 0.19  & $-$2.16 \\
6341 & II-53    & SKPL  & 4370 & 0.90  & $-$2.11 & 1.93 &   9 & 5.42 & 0.07 &   3 & 5.37 & 0.10  & $-$2.10 \\
6341 & V-45     & SKPL  & 4530 & 1.22  & $-$2.21 & 1.83 &   7 & 5.31 & 0.08 &   3 & 5.34 & 0.05  & $-$2.21 \\
6341 & XII-8    & SKPL  & 4510 & 1.17  & $-$2.23 & 1.84 &   7 & 5.30 & 0.10 &   2 & 5.24 & 0.05  & $-$2.22 \\
6341 & III-65   & SKPL  & 4340 & 0.94  & $-$2.12 & 1.92 &   8 & 5.41 & 0.08 &   2 & 5.43 & 0.05  & $-$2.11 \\
6352 & 181      & G8689 & 4235 & 1.50  & $-$0.63 & 1.23 &  15 & 6.89 & 0.18 &     &      &       & $-$0.63 \\
6352 & 111      & G8689 & 4415 & 1.80  & $-$0.78 & 1.88 &  19 & 6.74 & 0.19 &     &      &       & $-$0.78 \\
6352 & 142      & G8689 & 4325 & 1.70  & $-$0.52 & 1.14 &  17 & 7.00 & 0.17 &     &      &       & $-$0.52 \\
6362 & 32       & G8689 & 4210 & 1.10  & $-$0.98 & 1.71 &  22 & 6.55 & 0.16 &     &      &       & $-$0.97 \\
6362 & 25       & G8689 & 4240 & 1.10  & $-$0.96 & 1.74 &  22 & 6.56 & 0.15 &     &      &       & $-$0.96 \\
6397 & C43      & G8689 & 4526 & 1.30  & $-$1.64 & 1.45 &  11 & 5.88 & 0.15 &   3 & 6.19 & 0.09  & $-$1.64 \\
6397 & 302      & M93   & 4400 & 1.00  & $-$1.74 & 1.74 &  27 & 5.78 & 0.15 &   3 & 5.52 & 0.17  & $-$1.74 \\
6397 & C211     & NDC   & 4200 & 0.70  & $-$1.99 & 1.84 &  46 & 5.53 & 0.08 &   7 & 5.59 & 0.13  & $-$1.99 \\
6397 & C428     & G8689 & 4669 & 1.60  & $-$1.68 & 1.54 &   7 & 5.83 & 0.12 &   2 & 5.97 & 0.00  & $-$1.69 \\
6397 & 669      & M93   & 4421 & 1.00  & $-$1.87 & 1.71 &  25 & 5.65 & 0.11 &   2 & 5.64 & 0.07  & $-$1.87 \\
6397 & 603      & M93   & 4374 & 0.90  & $-$1.81 & 1.62 &  23 & 5.71 & 0.13 &   2 & 5.63 & 0.05  & $-$1.81 \\
6397 & 468      & M93   & 4600 & 1.50  & $-$2.04 & 1.83 &  20 & 5.49 & 0.10 &     &      &       & $-$2.03 \\
6397 & A331     & M93   & 4200 & 0.50  & $-$1.87 & 2.06 &  23 & 5.65 & 0.14 &     &      &       & $-$1.88 \\
6397 & C428     & CG95  & 4669 & 1.60  & $-$1.73 & 1.37 &  15 & 5.79 & 0.11 &   4 & 5.88 & 0.11  & $-$1.73 \\
6397 & C25      & CG95  & 4840 & 2.00  & $-$1.77 & 1.51 &   8 & 5.75 & 0.12 &   2 & 5.75 & 0.16  & $-$1.77 \\
6397 & C211     & CG95  & 4203 & 0.70  & $-$1.84 & 1.88 &  19 & 5.68 & 0.09 &   6 & 5.84 & 0.13  & $-$1.84 \\
6397 & 469      & NDC   & 4170 & 0.60  & $-$1.94 & 2.06 &  49 & 5.58 & 0.11 &   8 & 5.60 & 0.14  & $-$1.94 \\
6397 & C669     & G8689 & 4421 & 1.00  & $-$1.60 & 1.51 &  17 & 5.93 & 0.19 &   3 & 6.11 & 0.10  & $-$1.59 \\
6656 & III-52   & G8689 & 4192 & 0.80  & $-$1.41 & 1.90 &  25 & 6.12 & 0.15 &   5 & 6.20 & 0.15  & $-$1.40 \\
6656 & I-92     & G8689 & 4300 & 0.90  & $-$1.55 & 2.09 &  22 & 5.97 & 0.18 &   5 & 6.09 & 0.15  & $-$1.55 \\
6656 & III-12   & G8689 & 4189 & 0.80  & $-$1.49 & 2.17 &  24 & 6.03 & 0.12 &   3 & 6.16 & 0.16  & $-$1.49 \\
6752 & CL1015   & NDC   & 4350 & 1.10  & $-$1.46 & 1.57 &  55 & 6.06 & 0.14 &   8 & 6.12 & 0.15  & $-$1.46 \\
6752 & CS3      & NDC   & 4250 & 0.90  & $-$1.46 & 1.63 &  58 & 6.06 & 0.14 &   6 & 6.07 & 0.20  & $-$1.46 \\
6752 & CL1089   & NDC   & 4200 & 0.80  & $-$1.35 & 1.85 &  59 & 6.17 & 0.11 &   8 & 6.03 & 0.10  & $-$1.35 \\
6752 & CL1066   & NDC   & 4300 & 1.00  & $-$1.41 & 1.70 &  58 & 6.12 & 0.12 &   8 & 6.14 & 0.06  & $-$1.40 \\
6752 & A45      & NDC   & 4250 & 1.00  & $-$1.48 & 1.61 &  55 & 6.05 & 0.16 &   8 & 6.05 & 0.16  & $-$1.47 \\
6752 & 36       & M93   & 4400 & 1.00  & $-$1.32 & 1.58 &  29 & 6.20 & 0.11 &   3 & 6.10 & 0.10  & $-$1.32 \\
6752 & 284      & M93   & 4400 & 1.00  & $-$1.36 & 1.73 &  28 & 6.16 & 0.10 &   3 & 5.99 & 0.16  & $-$1.36 \\
6752 & A29      & NDC   & 4350 & 1.20  & $-$1.45 & 1.51 &  54 & 6.07 & 0.11 &   8 & 6.14 & 0.11  & $-$1.45 \\
6752 & 29       & M93   & 4350 & 1.20  & $-$1.35 & 1.33 &  24 & 6.17 & 0.10 &   2 & 6.37 & 0.13  & $-$1.35 \\
6752 & A31      & CG95  & 3915 & 0.30  & $-$1.48 & 2.25 &  27 & 6.05 & 0.10 &   4 & 6.17 & 0.09  & $-$1.47 \\
6752 & C3       & G8689 & 4260 & 0.90  & $-$1.34 & 1.41 &  24 & 6.17 & 0.16 &   4 & 6.19 & 0.18  & $-$1.35 \\
6752 & C9       & G8689 & 4500 & 1.50  & $-$1.41 & 1.05 &  21 & 6.10 & 0.18 &   2 & 6.06 & 0.08  & $-$1.42 \\
6752 & C118     & G8689 & 4460 & 1.40  & $-$1.33 & 1.49 &  21 & 6.20 & 0.15 &   2 & 6.11 & 0.31  & $-$1.32 \\
6752 & A45      & CG95  & 4258 & 1.00  & $-$1.41 & 1.63 &  23 & 6.11 & 0.11 &   5 & 6.17 & 0.08  & $-$1.41 \\
6752 & A61      & CG95  & 4310 & 1.10  & $-$1.61 & 2.04 &  20 & 5.91 & 0.10 &   5 & 6.12 & 0.09  & $-$1.61 \\
6752 & C9       & CG95  & 4498 & 1.50  & $-$1.48 & 1.89 &  15 & 6.04 & 0.14 &   3 & 6.04 & 0.07  & $-$1.48 \\
6809 & 283      & M93   & 4400 & 1.00  & $-$1.82 & 1.85 &  33 & 5.69 & 0.09 &   2 & 5.67 & 0.05  & $-$1.83 \\
6809 & 76       & M93   & 4400 & 1.00  & $-$2.10 & 1.88 &  26 & 5.42 & 0.11 &     &      &       & $-$2.10 \\
6838 & A4       & SKPL  & 4100 & 0.80  & $-$0.59 & 1.93 &  11 & 6.93 & 0.12 &   4 & 6.78 & 0.13  & $-$0.59 \\
6838 & I-45     & SKPL  & 4050 & 0.80  & $-$0.67 & 1.82 &  13 & 6.85 & 0.12 &   2 & 6.85 & 0.03  & $-$0.67 \\
6838 & I-113    & SKPL  & 3950 & 0.70  & $-$0.72 & 1.76 &  10 & 6.81 & 0.11 &   2 & 6.70 & 0.05  & $-$0.71 \\
6838 & I-46     & SKPL  & 4000 & 0.80  & $-$0.66 & 1.89 &  10 & 6.87 & 0.15 &   2 & 6.86 & 0.17  & $-$0.65 \\
6838 & I        & SKPL  & 4200 & 1.00  & $-$0.87 & 1.86 &  10 & 6.65 & 0.06 &   4 & 6.49 & 0.11  & $-$0.87 \\
6838 & I-53     & SKPL  & 4300 & 1.40  & $-$0.68 & 1.78 &  16 & 6.84 & 0.13 &   4 & 6.69 & 0.08  & $-$0.68 \\
6838 & I-77     & SKPL  & 4100 & 0.95  & $-$0.67 & 1.76 &  12 & 6.86 & 0.12 &   3 & 6.89 & 0.02  & $-$0.66 \\
6838 & A9       & SKPL  & 4200 & 1.20  & $-$0.83 & 1.88 &  15 & 6.70 & 0.13 &   6 & 6.64 & 0.15  & $-$0.82 \\
6838 & S        & SKPL  & 4300 & 1.25  & $-$0.69 & 1.92 &  16 & 6.83 & 0.09 &   5 & 6.86 & 0.10  & $-$0.69 \\
6838 & 53       & G8689 & 4200 & 1.60  & $-$0.77 & 1.71 &  16 & 6.75 & 0.09 &     &      &       & $-$0.77 \\
6838 & 56       & G8689 & 4580 & 2.10  & $-$0.51 & 1.84 &  19 & 7.01 & 0.09 &   2 & 6.68 & 0.19  & $-$0.51 \\
6838 & I-21     & SKPL  & 4350 & 1.45  & $-$0.70 & 1.93 &  16 & 6.82 & 0.15 &   5 & 6.80 & 0.10  & $-$0.70 \\
6838 & 21       & G8689 & 4400 & 1.65  & $-$0.78 & 1.51 &  16 & 6.73 & 0.11 &   1 & 6.78 &       & $-$0.79 \\
7078 & II-75    & M93   & 4416 & 0.80  & $-$2.10 & 1.65 &  24 & 5.41 & 0.16 &   1 & 5.07 &       & $-$2.11 \\
7078 & II-75    & SKPL  & 4410 & 0.90  & $-$2.11 & 1.93 &   8 & 5.41 & 0.12 &   2 & 5.03 & 0.12  & $-$2.11 \\
7078 & s6       & M93   & 4460 & 1.00  & $-$2.11 & 1.56 &  22 & 5.40 & 0.21 &   3 & 5.49 & 0.15  & $-$2.12 \\
7078 & IV-38    & SKPL  & 4300 & 0.60  & $-$2.13 & 2.03 &   8 & 5.38 & 0.20 &   1 & 5.26 &       & $-$2.14 \\
7078 & S1       & SKPL  & 4410 & 0.90  & $-$2.13 & 1.93 &   8 & 5.39 & 0.09 &   1 & 5.45 &       & $-$2.13 \\
7099 & D        & M93   & 4600 & 1.50  & $-$1.91 & 1.39 &  24 & 5.61 & 0.28 &   2 & 5.58 & 0.16  & $-$1.91 \\
7099 & 157      & M93   & 4600 & 1.50  & $-$1.90 & 1.30 &  21 & 5.62 & 0.17 &   3 & 5.32 & 0.11  & $-$1.90 \\
\\
\hline
\end{tabular}
\label{t:atmos}
}
\end{center}
\end{table*}


\begin{thebibliography}{}

\bibitem{} Anders, E., Grevesse, N. 1989, Geochim. Cosmochim. Acta, 53, 197
\bibitem{} Armandroff, T.E., Zinn, R. 1988, AJ, 96, 92
\bibitem{} Arp,  H.C. 1955, AJ, 60, 317
\bibitem{} Bard, A., Kock, A., Kock, M. 1991, A\&A, 248, 315
\bibitem{} Bard, A., Kock, M. 1994, A\&A, 282, 1014
\bibitem{} Bell, R.A., Gustafsson, B. 1989, MNRAS, 236, 653
\bibitem{} Bell, R.A., Eriksson, K., Gustafsson, B., Nordlund, A. 1976,
A\&AS, 23, 37 (BEGN)
\bibitem{} Bi\'emont, E., Baudoux, M., Kurucz, R.L., Ansbacher, W., Pinnington,
E.H. 1991, A\&A, 249, 539,
\bibitem{} Bingham, E.A., Cacciari, C., Dickens, R.J., Fusi Pecci, F. 1984, MNRAS, 209, 765
\bibitem{} Blackwell, D.E., Lynas-Gray, A.E. 1994, AA, 282, 899
\bibitem{} Blackwell, D.E., Shallis, M.J., Simmons, G.J. 1980, A\&A, 81, 340
\bibitem{} Butler, D., 1975, ApJ, 200, 68
\bibitem{} Carbon, D., Langer, G., Butler, D., Kraft, R., Suntzeff, N., Kemper,
E., Trefzger, C., Romanishin, W. 1982, ApJS, 49, 207
\bibitem{} Carretta, E., Gratton, R.G., Sneden, C. 1996, in preparation
\bibitem{} Castelli, F., Gratton, R.G., Kurucz, R.L. 1996, submitted to A\&A
\bibitem{} Clementini, G., Carretta, E., Gratton, R.G., Merighi, R., Mould, J.R.,
McCarthy, J.K. 1995, AJ, 110, 2319
\bibitem{} Cohen, J.G. 1983, ApJ, 270, 654 
\bibitem{} Cohen, J., Frogel, J.A., Persson, S.E. 1978, ApJ, 222, 165 (CFP)
\bibitem{} Costar, D., Smith, H.A. 1988, AJ, 96, 1925
\bibitem{} Dalle Ore, C. 1992, Ph.D. Thesis, Un. California
\bibitem{} Dalle Ore, C., Gratton, R.G., Peterson, R. 1996, in preparation
\bibitem{} Di Benedetto, G.P., Rabbia, Y. 1987, AA, 188, 114
\bibitem{} Djorgovski, S. 1993, {\it In} Structure and Dynamics of Globular
 Clusters, eds. S.G. Djorgovski and G. Meylan, ASP Conf. Ser. 50, p. 373
\bibitem{} Fernley, J. 1994, AA, 268, 591
\bibitem{} Frogel, J.A., Persson, S.E., Cohen, J.G. 1979, ApJ, 227, 499
\bibitem{} Frogel, J.A., Persson, S.E., Cohen, J.G. 1981, ApJ, 246, 842
\bibitem{} Frogel, J.A., Persson, S.E., Cohen, J.G. 1983, ApJS, 53, 713
\bibitem{} Gratton, R.G. 1987, A\&A, 179, 181 (G87)
\bibitem{} Gratton, R.G. 1988, Rome Obs. Preprint Ser. 29
\bibitem{} Gratton, R.G., Carretta, E., Castelli, f. 1996, submitted to A\&A
\bibitem{} Gratton, R.G., Ortolani, S. 1989, A\&A, 211, 41 (G89)
\bibitem{} Gratton, R.G., Sneden, C. 1994, A\&A, 287, 927
\bibitem{} Gratton, R.G., Quarta, M.L., Ortolani,S. 1986, A\&A, 169, 208 (G86)
\bibitem{} Hannaford, P., Lowe, R.M., Grevesse, N., Noels, A. 1992, A\&A, 259,
301
\bibitem{} Heise, C., Kock, M. 1990, A\&A, 230, 244
\bibitem{} Holweger, H., M\"uller, E.A. 1974, Solar Phys., 39, 19
\bibitem{} Kraft, R.P. 1994, PASP, 106, 553
\bibitem{} Kraft, R.P., Sneden, C., Langer, G.E., Prosser, C.F. 1992, AJ, 104,
645 (SKPL2)
\bibitem{} Kraft, R.P., Sneden, C., Langer, G.E., Shetrone, M.D. 1993, AJ, 106,
1490 (SKPL4)
\bibitem{} Kraft, R.P., Sneden, C., Langer, G.E., Shetrone, M.D., Bolte, M. 
1995, AJ, 109, 2586 (SKPL6) 
\bibitem{} Kurucz, R.L., 1992, private communication (K92)
\bibitem{} Kurucz, R.L., 1979, ApJS, 40, 1
\bibitem{} Leep, E.M., Oke, J.B., Wallerstein, G. 1987, AJ, 93, 338
\bibitem{} M\"ackle, R., Griffin, R., Griffin, R., Holweger, H. 1975, A\&AS, 19,
303
\bibitem{} Magain, P. 1984, A\&A, 134, 189
\bibitem{} Manduca, A. 1981, ApJ, 245, 258
\bibitem{} McWilliam, A., Geisler, D., Rich, R.M. 1992, PASP, 104, 1193
\bibitem{} Minniti, D., Geisler, D., Peterson, R.C., Claria, J.J. 1993, ApJ,
 413, 548
\bibitem{} Norris, J., Da Costa, G. 1995, ApJ, 447, 680
\bibitem{} Oosterhoff, P. 1944, BAN, 10, 55
\bibitem{} Pilachowski, C., Sneden, C., Wallerstein, G. 1983, ApJS, 52, 241
\bibitem{} Preston, G.W. 1959, ApJ, 130, 507
\bibitem{} Rutten, R.J., van der Zalm, E.B.J. 1984, A\&AS, 55, 143
\bibitem{} Sandage, A. 1990, ApJ, 350, 603
\bibitem{} Sandage, A. 1993a, AJ, 106, 703
\bibitem{} Sandage, A. 1993b, AJ, 106, 719
\bibitem{} Simmons, G.J., Blackwell, D.E. 1982, A\&A, 112, 209
\bibitem{} Smith, H.A. 1984, ApJ, 281, 148
\bibitem{} Smith, H.A. 1987, PASP, 94, 122
\bibitem{} Sneden, C., Kraft, R.P., Prosser, C.F., Langer, G.E. 1991, AJ, 102,
2001 (SKPL1)
\bibitem{} Sneden, C., Kraft, R.P., Prosser, C.F., Langer, G.E. 1992, AJ, 104,
2121 (SKPL3)
\bibitem{} Sneden, C., Kraft, R.P., Langer, G.E., Prosser, C.F., Shetrone, M.D.
1994, AJ, 107, 1773 (SKPL5)
\bibitem{} Trefzger, C., Carbon, D., Langer, G., Suntzeff, N., Kraft, R. 1983,
ApJ, 266, 144
\bibitem{} Tucholke, H.-J. 1992, A\&AS, 93, 293
\bibitem{} van Albada, T.S., Baker, N. 1973, ApJ, 185, 477
\bibitem{} Zinn, R., 1985, ApJ, 293, 424
\bibitem{} Zinn, R. 1980, ApJS, 42, 19
\bibitem{} Zinn, R., West, M.J. 1984, ApJS, 55, 45 (ZW)
\end{thebibliography}
\end{document}